\documentclass[12pt]{article}

\usepackage[dvips]{graphicx}
\usepackage[fleqn]{amsmath}
\usepackage[varg]{txfonts}
\usepackage{latexsym}
\usepackage{bm}
\usepackage[usenames]{color}
\usepackage{amssymb,mathrsfs}
\usepackage{colortbl}
\usepackage{multicol}
\usepackage{algorithmic}
\usepackage[ruled]{algorithm2e}

\newtheorem{Def}{Def}[section]
\newtheorem{Defi}[Def]{Definition}
\newtheorem{Them}[Def]{Theorem}
\newtheorem{Lem}[Def]{Lemma}
\newtheorem{Cor}[Def]{Corollary}
\newtheorem{Prop}[Def]{Proposition}
\newtheorem{Examp}[Def]{Example}

\numberwithin{equation}{section}

\newcommand{\LM}{\operatorname{LM}}
\newcommand{\PRM}{\operatorname{PRM}}
\newcommand{\RM}{\operatorname{RM}}

\newcommand{\supp}{\operatorname{supp}}

\newcommand{\NB}{\mathbb{N}}

\newcommand{\FC}{\mathcal{F}}

\newcommand{\FB}{\mathbb{F}}
\newcommand{\AB}{\mathbb{A}}
\newcommand{\PB}{\mathbb{P}}
\newcommand{\GC}{\mathcal{G}}

\title{Decoding of Projective Reed--Muller Codes by Dividing\\
a Projective Space into Affine Spaces}
\author{Norihiro Nakashima\footnote{Email: nakashima@toyota-ti.ac.jp}, 
Hajime Matsui\footnote{Email: matsui@toyota-ti.ac.jp}\\
Toyota Technological Institute, Nagoya 468-8511, Japan.}
\date{}

\begin{document}

\maketitle

\begin{abstract}
A projective Reed--Muller (PRM) code, obtained by modifying
a Reed--Muller code with respect to a projective space,
is a doubly extended Reed--Solomon code when the dimension of the
related projective space is equal to $1$.
The minimum distance and the dual code of a PRM code
are known, and some decoding examples have been presented for
low-dimensional projective spaces.
In this study, we construct a decoding algorithm for all PRM codes
by dividing a projective space into a union of affine spaces.
In addition, we determine the computational complexity and
the number of errors correctable of our algorithm.
Finally, we compare the codeword error rate of our algorithm with that of
the minimum distance decoding.
\vspace{5mm}
\vspace{2mm}
\end{abstract}

\section{Introduction}
Projective Reed--Muller (PRM) codes have been investigated extensively
since they were first introduced by Lachaud \cite{Lachaud} in 1988.
S{\o}rensen \cite{Sorensen} determined the minimum distances of PRM codes
and proved that the dual code of a PRM code is also a PRM code
or is spanned by a PRM code and a vector of ones.
In addition, Berger and Maximy \cite{Berger-Maximy} presented
conditions under which PRM codes are cyclic or quasi-cyclic.
Recently, Ballet and Rolland \cite{Ballet-Rolland}
examined low-weight codewords of PRM codes
and obtained an estimation of the second weight.
The PRM codes of one-dimensional projective spaces are also
considered to be doubly extended Reed--Solomon codes.
Decoding examples for PRM codes related to
low dimensional projective spaces
are presented in \cite{Dur}, \cite{Duursma}, \cite{Nakashima-MatsuiISITA2014}.

To realize practical communication channels,
many researchers constructed decoding procedures
whose computational complexities are polynomial time.
In addition, they investigated the numbers
of errors correctable and the codeword error rates.
Although the minimum distance decoding (MDD)
\cite{Justesen}, \cite{MacWilliams-Slone}
achieves a good codeword error rate,
the computational complexity of the MDD
based on generating all codewords is known to be exponential.
Pellikaan \cite{Pellikaan} developed a decoding algorithm
for linear codes, which corrects $t$-errors
if there exist $t$-error correcting pairs.
The computational complexity of this algorithm is $O(n^3)$,
where $n$ is the code length.
The Feng--Rao decoding algorithm \cite{Feng-Rao1}, \cite{Miura}
is also shown as a decoding method of $O(n^3)$ for linear codes.
The number of errors correctable by the Feng--Rao algorithm
is determined by Feng--Rao bounds
\cite{Feng-Rao1}, \cite{Miura}, \cite{Andersen-Geil}, \cite{Matsumoto-Miura}.
It is possible that these two algorithms can be applied to PRM codes.
However, we cannot find any observations of $t$-error correcting pairs
and Feng--Rao bounds for PRM codes,
and it is difficult to determine the numbers of errors correctable.

The objective of the present study is to investigate a decoding algorithm
for all PRM codes such that its computational complexity
is less than $O(n^3)$ and the number of errors correctable is determined.
We construct a new decoding algorithm by dividing
a projective space into a union of affine spaces
that a decoding algorithm proposed the second author \cite{Matsuiaffine}
is applied for each affine component.
In our algorithm, we adopt the Berlekamp--Massey--Sakata (BMS) algorithm
\cite{Sakata1}, \cite{Sakata2}, \cite{Sakata3}, \cite{Sakata4}
to obtain a Gr{\"o}bner basis whose zeros are the error positions,
and we use the discrete Fourier transform (DFT) to determine the error values.
After that, we prove that the computational complexity of our algorithm is
strictly less than $O(n^3)$.
In particular, the complexity of the error position determination
is $O(zn^{2})$ and that of the error value determination
is $O(qn^{2})$, where $z$ is the maximum of the cardinalities
of Gr{\"o}bner bases obtained by BMS algorithm for all components and
$q$ is the finite field cardinality. We have $z<n/q$.
Next, we determine the number of errors correctable
by our algorithm component-wise.
This implies the number of errors correctable at arbitrary positions.
Finally, we compare the codeword error rate of our algorithm with
that of the MDD and find them to be similar for some high-order PRM codes.

The remainder of this paper is organized as follows.
In Section \ref{sec-Preliminaries}, we present some preliminary notation and
recall the results of a previous study \cite{Matsuiaffine}.
In Section \ref{Sec-RepAVC}, we present
an example of PRM code that shows a difficulty
to construct a decoding algorithm.
In Section \ref{Sec-DecAlg},
we construct a decoding algorithm for all PRM codes.
In Section \ref{Sec-number}, we determine the number
of errors corrected by our algorithm.
In Section \ref{sec-example}, we present an example of a decoding procedure.
In Section \ref{sec-complexity}, we compute
the computational complexity of our algorithm.
In Section \ref{Sec-comparison},
we compare the codeword error rate of our algorithm with that of MDD.
Finally, in Section \ref{sec-conclusion}, we summarize our findings
and conclude the paper by briefly discussing
the scope for future investigation.

\section{Preliminaries}\label{sec-Preliminaries}
\subsection{Reed--Muller codes}
Throughout this paper, let $q$ be a prime power and let $\FB_{q}$ denote
a finite field consisting of $q$ elements.
Let $m$ be a positive integer. We define
\begin{equation}
\AB_{m}(\FB_{q})=\left\{(\omega_{1},\dots,\omega_{m})\,\middle|\,
\omega_{1},\dots,\omega_{m}\in\FB_{q}\right\},
\end{equation}
where $\AB_{m}(\FB_{q})$ is called an $m$-dimensional
affine space over $\FB_{q}$.
We often omit a coefficient field $\FB_{q}$ and
write $\AB_{m}(\FB_{q})=\AB_{m}$ for short.
Let $\FB_{q}[X_{1},\dots,X_{m}]$ denote the polynomial ring
over $\FB_{q}$ in $m$ variables.
For a polynomial $f(X_{1},\dots,X_{m})\in\FB_{q}[X_{1},\dots,X_{m}]$,
we often write $f(X_{1},\dots,X_{m})=f$.
Let $f(\omega_{1},\dots,\omega_{m})$ denote
the value obtained by substituting $(\omega_{1},\dots,\omega_{m})\in\AB_{m}$
for $f\in\FB_{q}[X_{1},\dots,X_{m}]$.

Let $\FB_{q}[X_{1},\dots,X_{m}]_{\leq\nu}$
denote the set of all polynomials
in $\FB_{q}[X_{1},\dots,X_{m}]$ of degree $\leq\nu$.
\begin{Defi}[Reed--Muller code, RM code]
A RM code over $\FB_{q}$ of order $\nu$ and length $q^m$ is defined by
\begin{equation}\label{eq-def-RMcode}
\RM_{\nu}(m,q)
=\left\{(f(P))_{P\in\AB_{m}}\,\middle|\,
f\in \FB_{q}[X_{1},\dots,X_{m}]_{\leq\nu}\right\}.\ \ \Box
\end{equation}
\end{Defi}

It has been shown (cf.\ \cite{Blake-Mullin}) that the dimension $k$
and the minimum distance $d$ of $\RM_{\nu}(m,q)$ are
\begin{align}
&k=\sum_{t=0}^{\nu}\sum_{j=0}^m (-1)^j \binom{m}{j}
\binom{t-jq+m-1}{t-jq},\label{eq-dim-RM}\\
&d=(q-s)q^{m-r-1},\label{eq-MD-RM}
\end{align}
where $r$ and $s$ are respectively the quotient
and remainder obtained when
$\nu$ is divided by $q-1$; that is,
$0\leq r<m-1$, $0\leq s<q-1$, and $\nu=r(q-1)+s$.
For a finite set $\Omega$, let
$\FB_{q}^{\Omega}=\{(c_{P})_{P\in\Omega}\mid c_{P}\in\FB_{q}\}$
denote the $\FB_{q}$-linear space indexed by $\Omega$.
For a subset $C$ of $\FB_{q}^{\Omega}$,
we denote the dual $C^{\perp}$ of $C$ by
\begin{equation}
C^{\perp}=\left\{
(u_{P})_{P\in\Omega}\in\FB_{q}^{\Omega}\,\middle|\,
\sum_{P\in\Omega}c_{P}u_{P}=0\ {\rm for\ all}\ (c_{P})_{P\in\Omega}\in C
\right\}.
\end{equation}
The following is widely known (see, e.g., \cite{Sorensen}).
\begin{Prop}\label{Prop-RM-dual}
Let $\mu=m(q-1)-\nu$. The dual of $\RM_{\nu}(m,q)$ is obtained by
\begin{align}\label{eq-dual-RM}
\RM_{\nu}(m,q)^{\perp}=\RM_{\mu-1}(m,q).\qquad\Box
\end{align}
\end{Prop}

\subsection{Projective Reed--Muller codes}
We define
\begin{equation}
\PB_{m}(\FB_{q})=(\AB_{m+1}\setminus\{0\})/\sim
\end{equation}
with the equivalence relation
\begin{equation}
P_{1}\sim P_{2}\quad
{\rm if\quad}P_{1}=\lambda P_{2}\
{\rm for\ some\ }\lambda\in\FB_{q}\setminus\{0\},
\end{equation}
where $\PB_{m}(\FB_{q})$ is called an $m$-dimensional projective
space over $\FB_{q}$.
We often write $\PB_{m}(\FB_{q})=\PB_{m}$.

We express the equivalence class of a representative
$(\omega_{0},\omega_{1},\dots,\omega_{m})$ as
$(\omega_{0}:\omega_{1}:\dots:\omega_{m})$.
For each $P=(\omega_{0}:\omega_{1}:\dots:\omega_{m})\in\PB_{m}$,
let $i$ be the smallest index such that $\omega_{i}\neq 0$.
Then, $(0,\dots,0,1,\omega_{i+1}^{\prime},\dots,\omega_{m}^{\prime})$ is
a representative of $P$, where
$\omega_{j}^{\prime}=\omega_{j}/\omega_{i}$ for $j>i$.
Let $R$ denote the polynomial ring $\FB_{q}[X_{0},X_{1},\dots,X_{m}]$
over $\FB_{q}$ in variables $X_{0},X_{1},\dots,X_{m}$.
The value $f(P)$ is defined by substituting the representative
$(0,\dots,0,1,\omega_{i+1}^{\prime},\dots,\omega_{m}^{\prime})$
for $f=f(X_{0},X_{1},\dots,X_{m})\in R$; this is uniquely determined.
A projective space is identified by
a union of affine spaces, i.e.,
\begin{equation}
\PB_{m}=\Psi_{0}\cup\Psi_{1}\cup\cdots\cup\Psi_{m},
\end{equation}
where $\Psi_{i}=\{(0:\cdots:0:1:\omega_{i+1}:\cdots:\omega_{m})\in\PB_{m}
\mid \omega_{j}\in\FB_{q},\,i+1\leq j\leq m\}$
is a subset of $\PB_{m}$ for all $i\in\{0,1,\dots,m\}$ by which
an $(m-i)$-dimensional affine space is identified.

Let $n$ be the number of elements in $\PB_{m}$. Then,
$n=(q^{m+1}-1)/(q-1)=q^{m}+\cdots+q+1$.
Let $R_{\nu}$ denote the linear subspace of $R$ consisting
of homogeneous polynomials of degree $\nu$.
\begin{Defi}[Projective Reed--Muller code, PRM code]
A PRM code over $\FB_{q}$ of order $\nu$
and length $n$ is defined by
\begin{equation}\label{def-PRM}
\PRM_{\nu}(m,q)=\left\{(f(P)_{P\in\PB_{m}}\,\middle|\,
f\in R_{\nu}\right\}.\quad\Box
\end{equation}
\end{Defi}

A PRM code is trivial (i.e., $\dim\PRM_{\nu}(m,q)=n$)
if $\nu>m(q-1)$ (see \cite[Remark 3]{Sorensen}).
Therefore, in the rest of this paper, we assume that $0<\nu\leq m(q-1)$.
It is shown (cf.\ \cite{Sorensen}) that $\PRM_{\nu}(m,q)$
is an $(n, k, d)$-code with
\begin{align}
&k=\sum_{t=0}^{r}\left(\sum_{j=0}^{m+1} (-1)^j \binom{m+1}{j}
\binom{s+m-t+(t-j)q}{s+1-t+(t-j)q}\right),\\
&d=(q-s)q^{m-r-1},
\end{align}
where $r$ and $s$ are determined by
$0\leq r<m$, $0\leq s<q-1$, and $\nu-1=r(q-1)+s$.
Table \ref{Table-parameter-PRMcodes} lists
some dimensions and minimum distances of $\PRM_{\nu}(2,16)$.
The following is used later in Lemma \ref{lem-synd-PRMkaraRM}.
\begin{Them}[\cite{Sorensen}]\label{PRMperp}
Let $\mu=m(q-1)-\nu$.
The dual of $\PRM_{\nu}(m,q)$ is obtained by the following:
\begin{enumerate}
\item $\PRM_{\nu}(m,q)^{\perp}=\PRM_{\mu}(m,q)$ if
$\nu\not\equiv 0\ ({\rm mod}\ q-1)$,
\item $\PRM_{\nu}(m,q)^{\perp}=
{\rm span}_{\FB_{q}}\{\bm{1},\PRM_{\mu}(m,q)\}$
if\ $\nu\equiv 0\ ({\rm mod}\ q-1)$,\
where\ $\bm{1}=(1,\dots,1)\in\FB_{q}^{n}$.\hfill$\Box$
\end{enumerate}
\end{Them}
\begin{table}[t]
{\footnotesize
\caption{Parameters of $\PRM_{\nu}(2,16)$}
\label{Table-parameter-PRMcodes}
\begin{center}
\begin{tabular}{|c||c|c|c|c|c|c|c|c|c|}
\hline
 $\nu$ &$5$&$8$&$11$&$14$&$17$&$20$&$23$&$26$&$29$\\
\hline
 $k$ &$21$&$45$&$78$&$120$&$168$&$207$&$237$&$258$&$270$\\
\hline
 $d$ &$192$&$144$&$96$&$48$&$15$&$12$&$9$&$6$&$3$\\
\hline
\end{tabular}
\end{center}
}
\end{table}

\subsection{Affine variety codes}\label{subsec-AVC}
Let $\Psi$ be a non-empty subset of $\AB_{m}$, i.e.,
$\emptyset\neq\Psi\subseteq\AB_{m}$.
We define an ideal $Z(\Psi)$ of $\FB_{q}[X_{1},\dots,X_{m}]$ as
\begin{equation}
Z(\Psi)=\{f\in \FB_{q}[X_{1},\dots,X_{m}]
\mid f(P)=0\ {\rm for\ all}\ P\in\Psi\}.
\end{equation}
\begin{Defi}[Affine variety code]
For an $\FB_q$-linear subspace $L$
of a quotient ring\\
$\FB_{q}[X_{1},\dots,X_{m}]/Z(\Psi)$,
we define an affine variety code as
\begin{equation}\label{affi-var-code}
C(L,\Psi)=
\{(f(P))_{P\in\Psi}\in\FB_{q}^{\Psi}\mid f\in L\}.\quad\Box
\end{equation}
\end{Defi}

We previously proposed a decoding algorithm \cite[Algorithm 2]{Matsuiaffine}
for a class of affine variety codes using the BMS algorithm and DFT.
The following definitions are required to explain this decoding algorithm.
Let $M$ be the set of all monomials
whose exponent of each variable is less than $q$, i.e.,
$M=\{X_{1}^{a_{1}}\cdots X_{m}^{a_{m}}\mid
(a_{1},\dots,a_{m})\in\NB_{0}^m,\ a_{1},\dots,a_{m}\leq q-1\}$,
where $\NB_{0}$ is the set of nonnegative integers.
\begin{Defi}[Discrete Fourier transform, DFT]
A linear \\
map $\FC$ is defined by
\begin{equation}
\FC:\FB_{q}^{\AB_m}\rightarrow \FB_{q}^{M},\quad
(c_{P})_{P\in\AB_m}\mapsto\left(\sum_{P\in\AB_m}c_{P}h(P)\right)_{h\in M},
\end{equation}
and $\FC$ is called a DFT on $\FB_{q}^{\AB_m}$.\hfill$\Box$
\end{Defi}

The following map is the inverse of $\FC$, and
is called an inverse discrete Fourier transform (IDFT) on $\FB_{q}^{\AB_m}$.
For a finite set $\Omega$,
let $|\Omega|$ denote the number of elements in $\Omega$.
\begin{Defi}
For each $P=(\omega_{1},\dots,\omega_{m})\in\AB_m$,
we define a subset $\supp(P)$
of $\{1,\dots,m\}$ by $\supp(P)=\{i\mid \omega_{i}\neq 0\ (1\le i\le m)\}$.
Let $s=|\supp(P)|$. A linear map $\FC^{-1}$ is defined by
\begin{equation}
\FC^{-1}:\FB_{q}^{M}\rightarrow \FB_{q}^{\AB_m},\quad
(r_{h})_{h\in M}\mapsto (c_{P})_{P\in\AB_m},
\end{equation}
where
\begin{align}
c_{P}=(-1)^{s}\sum_{l_{1}=1}^{q-1}\cdots\sum_{l_{s}=1}^{q-1}
\left\{\sum_{J\subseteq \supp(P)^{c}}
(-1)^{|J|}r_{h_{(P,l,J)}}\right\}
\omega_{1}^{-l_{1}}\cdots \omega_{s}^{-l_{s}},
\end{align}
$J$ runs over all subsets of $\supp(P)^{c}=\{1,\dots,m\}\setminus\supp(P)$,
and $h_{(P,l,J)}=X_{1}^{b_{1}}\cdots X_{m}^{b_{m}}$ is a monomial such that
\begin{equation}
b_{i}=
\begin{cases}
l_{i}&\ {\rm if}\ i\in \supp(P),\\
q-1&\ {\rm if}\ i\in J,\\
0&\ {\rm if}\ i\not\in \supp(P)\cup J.\quad\Box
\end{cases}
\end{equation}
\end{Defi}

Let $\prec$ be a monomial order, and $\mathcal{G}_{\Psi}$
a Gr\"obner basis for the ideal $Z(\Psi)$
(see \cite{Cox1}, \cite{Cox2}, \cite{Saints-Heegaard}
or \cite{Fitzgerald-Lax} for
the theory of Gr\"obner bases).
We write $X^{\bm{a}}=X_{1}^{a_{1}}\cdots X_{m}^{a_{m}}$ for
$\bm{a}=(a_{1},\dots,a_{m})\in \NB_{0}^{m+1}$.
Let $f\in \FB_{q}[X_{1},\dots,X_{m}]$, where
$f=\sum_{\bm{a}\in \NB_{0}^{m}}\lambda_{\bm{a}}X^{\bm{a}}$
for some coefficients $\lambda_{\bm{a}}\in\FB_{q}$.
The leading monomial $\LM(f)$ of $f$
is the maximum of the monomials arranged in $\prec$
that have nonzero coefficients in $f$, i.e.,
$\LM(f)=\max_{\prec}\{X^{\bm{a}}\mid \lambda_{\bm{a}}\neq 0\}$.
For a subset $\Phi$ of $\Psi$, we define a set $D(\Phi)$ as
\begin{equation}\label{eq-def-of-Delta}
D(\Phi)=\{X^{\bm{a}}\mid\bm{a}\in\NB_{0}^{m}\}
\setminus\{\LM(f)\mid 0\neq f\in Z(\Phi)\}.
\end{equation}
Since $\{X_{1}^q-X_{1},\dots,X_{m}^q-X_{m}\}\subseteq Z(\Psi)$,
we have $D(\Phi)\subseteq D(\Psi)\subseteq M$.
We note that $D(\Psi)$ forms a basis for $\FB_{q}[X_{1},\dots,X_{m}]/Z(\Psi)$
(see \cite[Theorem 19]{Saints-Heegaard}).

Let $z$ be the number of elements in the Gr\"obner basis $\mathcal{G}_{\Phi}$,
and $\{f^{(1)},\dots,f^{(z)}\}$ the set of elements in $\mathcal{G}_{\Phi}$.
\begin{Defi}
A linear map $\mathcal{E}_{\Phi}$ is defined by
\begin{equation}
\mathcal{E}_{\Phi}:\FB_{q}^{D(\Phi)}\rightarrow \FB_{q}^{M},\quad
(r_{h})_{h\in D(\Phi)}\mapsto (r_{g})_{g\in M},
\end{equation}
where for $g\in M$,
\begin{equation}
r_{g}=\sum_{h\in D(\Phi)}v_{h}r_{h},
\end{equation}
$v_{h}$ is obtained by the division algorithm by $\mathcal{G}_{\Phi}$:
\begin{equation}
g(X)=\sum_{0\leq w<z}u^{(w)}(X)f^{(w)}(X)+v(X)
\end{equation}
for some $u^{(w)}(X)\in \FB_{q}[X_{1},\dots,X_{m}]$ and
$v(X)=\sum_{h\in D(\Phi)}v_{h}h\in \FB_{q}[X_{1},\dots,X_{m}]$.\hfill$\Box$
\end{Defi}
\begin{Defi}
Let $L$ be a subspace of $\FB_{q}[X_{1},\dots,X_{m}]/Z(\Psi)$
over $\FB_{q}$. We say that $L$ has a monomial basis if 
\begin{equation}
L={\rm span}_{\FB_{q}}(B)\ {\rm for\ some}\ 
B\subseteq D(\Psi).\quad\Box\label{eq-condetion-affinealgo}
\end{equation}
\end{Defi}
\begin{Examp}
Let $\Psi=\AB_{1}(\FB_{4})$. Then, $Z(\Psi)=\langle X^4+X\rangle$
and $D(\Psi)=\{1,X,X^2,X^3\}$. The linear space
$L={\rm span}_{\FB_{4}}\{1,X,X^2,X^3\}$ has
a monomial basis $B=\{1,X,X^2,X^3\}\subseteq D(\Psi)$.
Next, $L^{\prime}={\rm span}_{\FB_{4}}\{1+X^2\}$ does not
have any monomial basis, since $1+X^2$ is not in $D(\Psi)$.\hfill$\Box$
\end{Examp}
\begin{Examp}
Let $\Psi=\AB_{2}(\FB_{4})$.
Since $Z(\Psi)=\langle X_{1}^4+X_{1},X_{2}^4+X_{2}\rangle$, we have
$D(\Psi)=\{X_{1}^i X_{2}^j\mid 0\leq i,j\leq 3\}$.
Then, $L={\rm span}_{\FB_{4}}\{1,X_{1}+X_{2},X_{2}\}$
has a monomial basis $B=\{1,X_{1},X_{2}\}$,
since $X_{1}$ is a linear combination of $X_{1}+X_{2}$ and $X_{2}$.\hfill$\Box$
\end{Examp}
\begin{Examp}\label{example-RM-mono-basis}
Let $\Psi=\AB_{m}$. Then,
$Z(\Psi)=\langle X_{1}^q-X_{1},\dots X_{m}^q-X_{m}\rangle$.
We have that $C(L,\Psi)=\RM_{\nu}(m,q)$, where
$B=\{\prod_{j=1}^{m}X_{j}^{a_{j}}\mid \sum_{j=1}^{m}a_{j}\leq \nu,\ 
0\leq a_{1},\dots,a_{m}\leq q-1\}$ and $L={\rm span}_{\FB_{q}}(B)$.
Thus, $L$ has a monomial basis $B$.\hfill$\Box$
\end{Examp}

Let $(r_{P})_{P\in\Psi}=(c_{P})_{P\in\Psi}+(e_{P})_{P\in\Psi}$
be a received word, where $(c_{P})_{P\in\Psi}\in C^{\perp}(L,\Psi)$
and $(e_{P})_{P\in\Psi}\in\FB_{q}^{\Psi}$.
Let $\Phi=\{P\in\Psi\mid e_{P}\neq 0\}$ be the set of error positions
of the received word $(r_{P})_{P\in\Psi}$.
We call $\left(\sum_{P\in\Psi}r_{P}h(P)\right)_{h\in B}$
a syndrome of $(r_{P})_{P\in\Psi}$ related to $C(L,\Psi)$.
It follows from $(c_{P})_{P\in\Psi}\in C^{\perp}(L,\Psi)$ that
$\left(\sum_{P\in\Psi}r_{P}h(P)\right)_{h\in B}
=\left(\sum_{P\in\Psi}e_{P}h(P)\right)_{h\in B}$.
Thus, the syndrome is a $B$-component of $\FC((e_{P})_{P\in\AB_{m}})$,
where $e_{P}=0$ if $P\in\AB_{m}\setminus\Psi$.
Let $\mathcal{R}_{\Psi}:\FB_{q}^{\AB_m}\rightarrow\FB_{q}^{\Psi}$
be the restriction map.
Algorithm \ref{AffineAlgo} is a decoding algorithm for $C^{\perp}(L,\Psi)$.
To apply Algorithm \ref{AffineAlgo}, it is sufficient
that $L$ has a monomial basis $B$.
We note that a RM code is expressed as $C^{\perp}(L,\Psi)$ such that
$L$ has a monomial basis by Proposition \ref{Prop-RM-dual}
and Example \ref{example-RM-mono-basis}.

\begin{algorithm}[h]
\caption{Error correction for $C^{\perp}(L,\Psi)$ \cite{Matsuiaffine}}
\label{AffineAlgo}
{\fontsize{10}{10}\selectfont
\KwIn{$(r_{P})_{P\in\Psi}\in \FB_{q}^{\Psi}$, where $(r_{P})_{P\in\Psi}=(c_{P})_{P\in\Psi}+(e_{P})_{P\in\Psi}$, $(c_{P})_{P\in\Psi}\in C^{\perp}(L,\Psi)$ and $(e_{P})_{P\in\Psi}\in \FB_{q}^{\Psi}$}
\KwOut{$(\hat{c}_{P})_{P\in\Psi}$}
\vspace{1mm}

Step 1.\ $(S_{h})_{h\in B}
=\left(\sum_{P\in\Psi}r_{P}h(P)\right)_{h\in B}$.
\vspace{1mm}

Step 2.\ Calculate $\mathcal{G}_{\Phi}$ from the syndrome
$(S_{h})_{h\in B}$ by \\
\quad\qquad\,the BMS algorithm
(cf.\ \cite{Cox2}, \cite{Amoros-Sullivan}).
\vspace{1mm}

Step 3.\ $(\hat{e}_{P})_{P\in\Psi}
=\mathcal{R}_{\Psi}\circ\mathcal{F}^{-1}\circ\mathcal{E}_{\Phi}
\left((S_{h})_{h\in B}\right)$.
\vspace{1mm}

Step 4.\ $(\hat{c}_{P})_{P\in\Psi}=
(r_{P})_{P\in\Psi}-(\hat{e}_{P})_{P\in\Psi}$.
}
\end{algorithm}

In the case when the dimension of $C^{\perp}(L,\Psi)$ is not $0$,
Algorithm \ref{AffineAlgo} computes $(c_{P})_{P\in\Psi}$ correctly,
i.e., $(\hat{c}_{P})_{P\in\Psi}=(c_{P})_{P\in\Psi}$, if
\begin{equation}\label{eq-FR-error}
2|\Phi|<d_{{\rm FR}}(C^{\perp}(L,\Psi)),
\end{equation}
where $d_{{\rm FR}}(C^{\perp}(L,\Psi))$ is a Feng--Rao bound.
In Step 1, we calculate a syndrome $(S_{h})_{h\in B}$ of $(r_{P})_{P\in\Psi}$.
In Step 2, we calculate the Gr{\"o}bner basis $\mathcal{G}_{\Phi}$
for $Z(\Phi)$ whose zeros are error positions.
In Step 3, we extend the syndrome 
$(S_{h})_{h\in B}=\left(\sum_{P\in\Psi}e_{P}h(P)\right)_{h\in B}$
to $\mathcal{F}((e_{P})_{P\in\AB_{m}})$ by applying $\mathcal{E}_{\Phi}$.
Then, by applying $\mathcal{R}_{\Psi}\circ\mathcal{F}^{-1}$,
we obtain the error word $(e_{P})_{P\in\Psi}$.

If the dimension of $C^{\perp}(L,\Psi)$ is $0$,
Algorithm \ref{AffineAlgo} computes all error words correctly, i.e.,
$(\hat{c}_{P})_{P\in\Psi}=(c_{P})_{P\in\Psi}$ for all
$(e_{P})_{P\in\Psi}\in\FB_{q}^{\Psi}$.
Indeed, since $L$ has a monomial basis $B=M$, we have
$(S_{h})_{h\in B}=(S_{h})_{h\in M}
=\left(\sum_{P\in\Psi}e_{P}h(P)\right)_{h\in M}
=\mathcal{F}\left((e_{P})_{P\in\AB_{m}}\right)$.
This means that the syndrome is the image of an error word by the DFT.
Thus, by applying $\mathcal{R}_{\Psi}\circ\mathcal{F}^{-1}$ to
the syndrome, we obtain the error word $(e_{P})_{P\in\Psi}$.
Hence, in this case, we do not calculate Step 2
and $\mathcal{E}_{\Phi}$ of Step 3.

\section{Basis for PRM codes}\label{Sec-RepAVC}
In general, if $L$ has a monomial basis and
a Feng--Rao bound of $C^{\perp}(L,\Psi)$ is high,
Algorithm \ref{AffineAlgo} has a good codeword error rate.
However, when $C^{\perp}(L,\Psi)$ is a PRM code,
it is difficult to determine whether $L$ has a monomial basis.
In this section, we present an example of PRM code $C^{\perp}(L,\Psi)$
such that $L$ does not have any monomial bases.

First, we prove that a PRM code is the dual of an affine variety code.
A projective space $\PB_{m}$ is identified by a set
$\Psi=\bigcup_{i=0}^{m}\{(0,\dots,0,1,\omega_{i+1},\dots,\omega_{n})\mid
\omega_{i+1},\dots,\omega_{n}\in\FB_{q}\}$
of representatives in $\AB_{m+1}$.
Let $\nu$ be a positive integer and $\mu=m(q-1)-\nu$.
Let $L={\rm span}_{\FB_{q}}\{X^{\bm{a}}\in R/Z(\Psi)
\mid \bm{a}\in\NB^{m+1}_{0},|\bm{a}|=\mu\}$
if $\nu\not\equiv 0$ modulo $q-1$, and
$L={\rm span}_{\FB_{q}}\{\bm{1},\ X^{\bm{a}}\in R/Z(\Psi)
\mid \bm{a}\in\NB^{m+1}_{0},|\bm{a}|=\mu\}$
if $\nu\equiv 0$ modulo $q-1$.
Then, $C^{\perp}(L,\Psi)=\PRM_{\nu}(m,q)$ by Eq.\ \eqref{def-PRM},
Eq.\ \eqref{affi-var-code} and Theorem \ref{PRMperp}.
To determine whether $L$ has a monomial basis, we need to consider
reductions in $R/Z(\Psi)$ and linear combinations of elements in $L$.

Next, we present an example of a PRM code
such that $L$ does not have any monomial bases.
Let $|\bm{a}|=a_{0}+a_{1}+\cdots+a_{m}$ for
$\bm{a}=(a_{0},a_{1},\dots,a_{m})\in \NB_{0}^{m+1}$.
In this section, we fix a monomial order $\prec$ in the following manner:
$X^{\bm{a}}\prec X^{\bm{b}}$ if ``$|\bm{a}|<|\bm{b}|$'' or
``$|\bm{a}|=|\bm{b}|$ and
there exists an index $\ell$ such that
$a_{m}=b_{m}, a_{m-1}=b_{m-1},\dots,a_{\ell+1}=b_{\ell+1}\
{\rm and}\ a_{\ell}<b_{\ell}$.''
\begin{Defi}
A set of polynomials $\GC$ is defined as follows:
\begin{enumerate}
\item When $m=1$, we set
$\GC=\{X_{1}^q-X_{1},(X_{0}-1)(X_{1}-1),X_{0}^2-X_{0}\}$.
\item When $m=2$, we set
$\GC=\{X_{2}^q-X_{2},X_{1}^q-X_{1},
(X_{0}-1)(X_{1}-1)(X_{2}-1),
(X_{0}-1)(X_{1}^2-X_{1}),
X_{0}^2-X_{0}\}.$\hfill$\Box$
\end{enumerate}
\end{Defi}

\begin{figure}[t]
\begin{center}
\includegraphics[width=50mm]{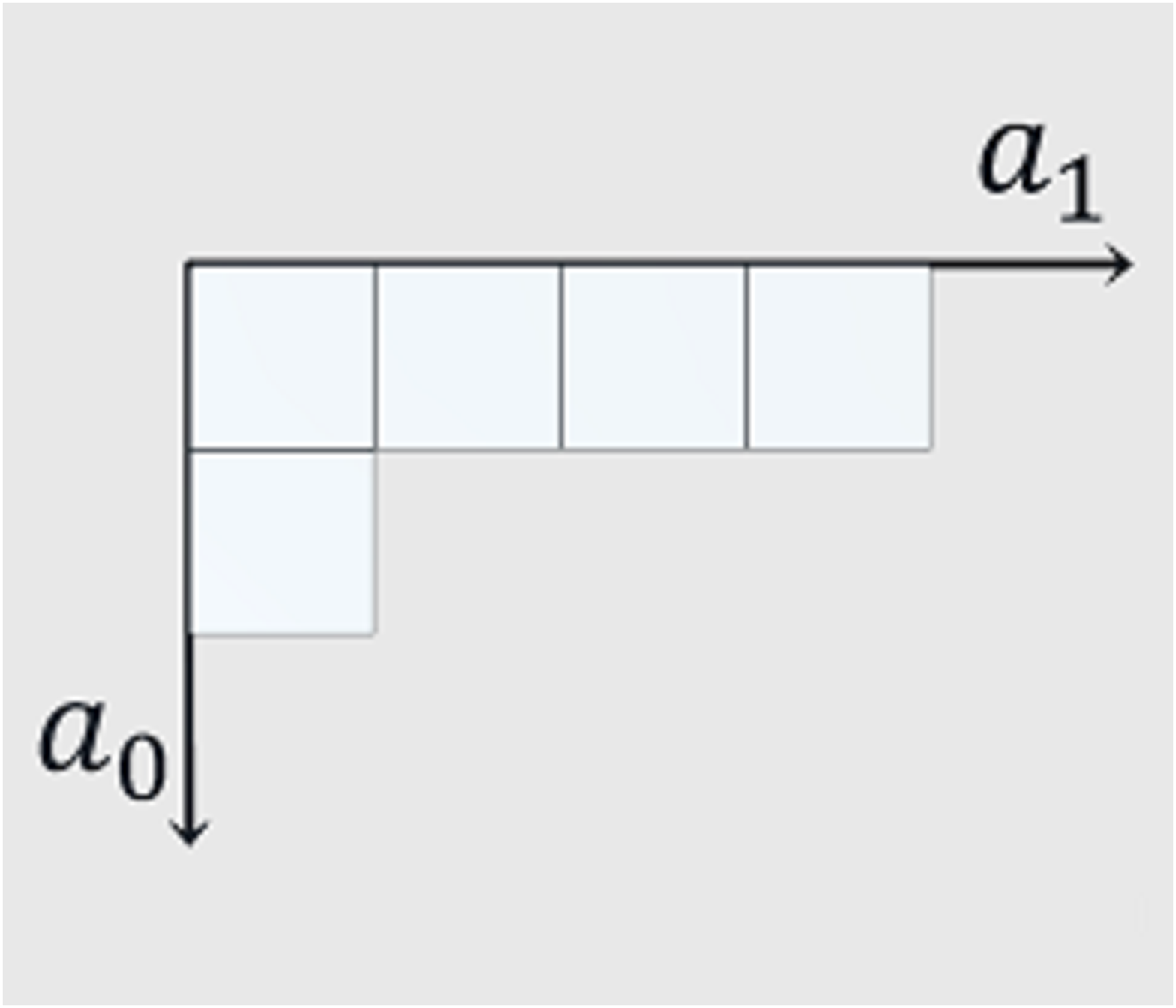}
\end{center}
 \caption{Monomial positions of $D(\Psi)$ if $m=1,q=4$}
 \label{fig-monobasis2}
\vspace{2mm}
\begin{center}
\includegraphics[width=50mm]{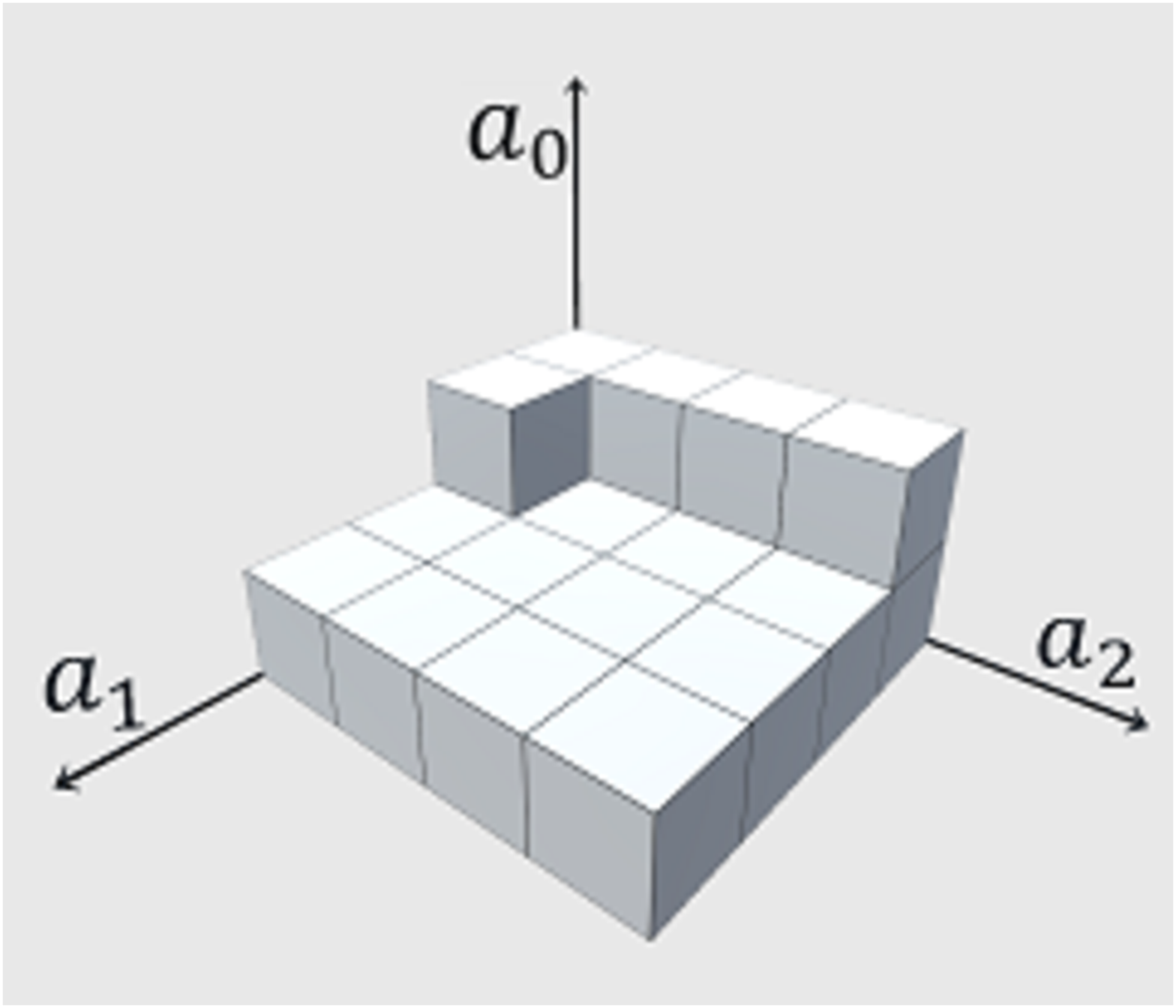}
\end{center}
 \caption{Monomial positions of $D(\Psi)$ if $m=2,q=4$}
 \label{fig-monobasis1}
\end{figure}

The inclusion $\GC\subseteq Z(\Psi)$ immediately follows.
Let $\langle\GC\rangle$ denote the ideal of $R$ generated by $\GC$.
By Buchberger's criterion (see \cite[Theorem 2.6.6]{Cox1}),
we can directly verify that $\GC$ is a Gr\"obner basis for $\langle\GC\rangle$.
Thus, we can compute a basis for a quotient ring $R/\langle\GC\rangle$,
and we have $\dim_{\FB_{q}}(R/\langle\GC\rangle)=n$
by \cite[Proposition 5.3.4]{Cox1}.
At the same time, we have $\dim_{\FB_{q}}(R/Z(\Psi))=|\Psi|=n$
by \cite[Theorem 19]{Saints-Heegaard}.
Therefore, $Z(\Psi)$ coincides with $\langle\GC\rangle$.
In particular, $\GC$ is a Gr\"obner basis for $Z(\Psi)$.
By \eqref{eq-def-of-Delta}, we have that
\begin{enumerate}
\item $D(\Psi)=\{X_{1}^{a_{1}}\mid 0\leq a_{1}\leq q-1\}\cup\{X_{0}\}$
if $m=1$,
\item $D(\Psi)=\{X_{1}^{a_{1}} X_{2}^{a_{2}}\mid 0\leq a_{1},a_{2}\leq q-1\}
\cup\{X_{0}X_{2}^{a_{2}}\mid 0\leq a_{2}\leq q-1\}\cup\{X_{0}X_{1}\}$
if $m=2$.
\end{enumerate}
We show monomial positions of $D(\Psi)$ in Fig.\ \ref{fig-monobasis2}
and Fig.\ \ref{fig-monobasis1}.
\begin{Examp}
Let $q=4,m=2,\nu=3$.
By Theorem \ref{PRMperp}, we have $C^{\perp}(L,\Psi)=\PRM_{3}(2,4)$, where
$L={\rm span}_{\FB_{4}}\{1,\ X^{\bm{a}}\mid |\bm{a}|=3\}
\subseteq \FB_{4}[X_{0},X_{1},X_{2}]/Z(\Psi)$.
Monomials $X_{0}X_{1}^2$, $X_{0}^2 X_{1}$
can be reduced in $\FB_{4}[X_{0},X_{1},X_{2}]/Z(\Psi)$ as follows:
\begin{align}
X_{0} X_{1}^2=X_{1}^2+X_{0}X_{1}-X_{1},\quad
X_{0}^2 X_{1}=X_{0}X_{1}.
\end{align}
Thus, $X_{1}^2-X_{1}$ is obtained by a linear combination of elements in $L$.
However, it follows from a direct calculation that
any linear combination of elements in $L$ containing $X_{1}^2-X_{1}$
is not in $D(\Psi)$.
This means that $L$ does not have any monomial bases.\hfill$\Box$
\end{Examp}

\section{Decoding algorithm}\label{Sec-DecAlg}
In this section, we construct a decoding algorithm
for all PRM codes following
the decomposition $\PB_{m}=\bigcup_{i=0}^{m}\Psi_{i}$.
As described in Section \ref{Sec-RepAVC},
there exists a PRM code that does not have any monomial bases.
On the other hand, for each $\Psi_{i}$-component,
we can find a suitable monomial basis $B_{i}$
such that $C^{\perp}({\rm span}_{\FB_{q}}(B_{i}),\Psi_{i})$ is a RM code.
Then, we obtain a $\Psi_{i}$-component
of an error ward $(e_{P})_{P\in\PB_{m}}$ by applying
Algorithm \ref{AffineAlgo} from a syndrome related to the RM code.
By repeating this for all $i\in\{0,1,\dots,m\}$,
we obtain the error word $(e_{P})_{P\in\PB_{m}}$.
We describe a non-trivial procedure to calculate the syndrome
in Lemma \ref{lem-synd-PRMkaraRM}.

Let $\nu$ be an integer where $0<\nu\leq m(q-1)$,
and let $\mu=m(q-1)-\nu$.
Let $(c_{P})_{P\in\PB_m}$ be a codeword in $\PRM_{\nu}(m,q)$.
After an error word $(e_{P})_{P\in\PB_m}$ occurs,
we assume that we receive the word
$(r_{P})_{P\in\PB_m}=(c_{P})_{P\in\PB_m}+(e_{P})_{P\in\PB_m}$.
Using the following settings, we can construct
a decoding algorithm by which
the error word $(e_{P})_{P\in\PB_m}$ may be corrected.

Let $i\in\{0,1,\dots,m\}$. 
We define a subset $B_{i}$ of $R_{\mu}$ by
\begin{equation}\label{eq-mono-Psi_i-before-homog}
B_{i}=\left\{\prod_{j=i}^{m}X_{j}^{a_{j}}\,\middle|\,
\begin{array}{c}
\sum_{j=i}^{m}a_{j}=\mu,\ 0<a_{i},\\
0\leq a_{i+1},\dots,a_{m}\leq q-1
\end{array}
\right\}.
\end{equation}
We recall that $\Psi_{i}$ is identified by
$\{(0,\dots,0,1,\omega_{i+1},\dots,\omega_{m})\mid
\omega_{i+1},\dots,\omega_{m}\in\FB_{q}\}\subseteq\AB_{m+1}$.
Since $Z(\Psi_{i})\subseteq R$ is generated by
$\{X_{0},\dots,X_{i-1},X_{i}-1,X_{i+1}^q-X_{i+1},\dots,X_{m}^q-X_{m}\}$,
we have $R/Z(\Psi_{i})=\FB_{q}[X_{i+1},\dots,X_{m}]/
\langle X_{i+1}^q-X_{i+1},\dots,X_{m}^q-X_{m}\rangle$.
Then, $B_{i}=\{\prod_{j=i+1}^{m}X_{j}^{a_{j}}\mid
\sum_{j=i+1}^{m}a_{j}\leq\mu-1,\ 0\leq a_{i+1},\dots,a_{m}\leq q-1\}$
in $R/Z(\Psi_{i})$, which is the set of monomials
in $D(\Psi_{i})$ of degree $\leq\mu-1$.
By Eq.\ \eqref{eq-def-RMcode} and Eq.\ \eqref{affi-var-code},
\begin{align}
C({\rm span}_{\FB_{q}}(B_{i}),\Psi_{i})=\RM_{\mu-1}(m-i,q).
\end{align}
Therefore, $\left(\sum_{P\in\Psi_{i}}e_{P}h(P)\right)_{h\in B_{i}}$
is a syndrome of $(e_{P})_{P\in\Psi_{i}}$
related to $\RM_{\mu-1}(m-i,q)$.
If we calculate the syndrome
$\left(\sum_{P\in\Psi_{i}}e_{P}h(P)\right)_{h\in B_{i}}$,
we can apply Step 2 and Step 3 of
Algorithm \ref{AffineAlgo} as $\Psi=\Psi_{i}$, $B=B_{i}$,
$(S_{h})_{h\in B}=\left(\sum_{P\in\Psi_{i}}e_{P}h(P)\right)_{h\in B_{i}}$ and
$C^{\perp}({\rm span}_{\FB_{q}}(B),\Psi)=\RM_{\mu-1}(m-i,q)^{\perp}$.
A procedure to obtain the syndrome is described later
in Lemma \ref{lem-synd-PRMkaraRM}.

\begin{algorithm}[h]
\caption{Decoding algorithm for $\PRM_{\nu}(m,q)$}
\label{decodingPRM}
{\fontsize{10}{10}\selectfont
\KwIn{$(r_{P})_{P\in\PB_{m}}\in\FB_{q}^{\PB_{m}}$, where $(r_{P})_{P\in\PB_m}=(c_{P})_{P\in\PB_m}+(e_{P})_{P\in\PB_m}$, $(c_{P})_{P\in\PB_m}\in\PRM_{\nu}(m,q)$ and $(e_{P})_{P\in\PB_m}\in\FB_{q}^{\PB_{m}}$}
\vspace{1mm}
\KwOut{$(\hat{e}_{P})_{P\in\PB_{m}}$}
\vspace{1mm}
\For{$i\in\{0,1,\dots,m\}$}{
\vspace{1mm}
(Step 1)\\
\vspace{1mm}
\eIf{$i=0$}{
\vspace{1mm}
$r^{(0)}_{P}=r_{P}$ for $P\in\PB_{m}$.}
{
\vspace{1mm}
$r^{(i)}_{P}=r_{P}-\hat{e}_{P}$ for
$P\in\bigcup_{j=0}^{i-1}\Psi_{j}$.\\
\vspace{1mm}
$r^{(i)}_{P}=r_{P}$ for
$P\in\bigcup_{j=i}^{m}\Psi_{j}$.
}
\vspace{1mm}
(Step 2)\\
\vspace{1mm}
Calculate $S^{(i)}_{h}=\sum_{P\in\PB_{m}}r^{(i)}_{P}h(P)$
for $h\in B_{i}$.\\
\vspace{1mm}
(Step 3)\\
\vspace{1mm}
Calculate $(\hat{e}_{P})_{P\in\Psi_{i}}$ by Algorithm \ref{AffineAlgo} as $\Psi=\Psi_{i}$, $B=B_{i}$ and $(S_{h})_{h\in B}=\left(S^{(i)}_{h}\right)_{h\in B_{i}}$.
}
}
\end{algorithm}

In Algorithm \ref{decodingPRM},
$(\hat{e}_{P})_{P\in\PB_{m}}=(e_{P})_{P\in\PB_{m}}$
if $(\hat{e}_{P})_{P\in\Psi_{i}}=(e_{P})_{P\in\Psi_{i}}$
for all $i\in\{0,1,\dots,m\}$.
Let $i_{0}$ be the smallest integer
satisfying $\mu\geq (m-i_{0})(q-1)+1$, i.e.,
\begin{equation}\label{eq-def-i_0}
i_{0}=m-\left\lfloor \frac{\mu-1}{q-1}\right\rfloor.
\end{equation}
If $i_{0}\leq i\leq m$, then
$(\hat{e}_{P})_{P\in\Psi_{i}}=(e_{P})_{P\in\Psi_{i}}$ for all
$(e_{P})_{P\in\Psi_{i}}\in\FB_{q}^{\Psi_{i}}$.
Indeed, since $\RM_{\mu-1}(m-i,q)=\FB_{q}^{\Psi_{i}}$,
the dimension of $\RM_{\mu-1}(m-i,q)^{\perp}$ is $0$
(see the last paragraph of Section \ref{subsec-AVC}).

Here, we explain how we obtain the syndrome
$\left(\sum_{P\in\Psi_{i}}e_{P}h(P)\right)_{h\in B_{i}}$
and how we apply Algorithm \ref{AffineAlgo} in Algorithm \ref{decodingPRM}.
We fix an integer $i$ where $0\leq i\leq m$.
In Step 1, if $i=0$, then we set $r^{(0)}_{P}=r_{P}$ for $P\in\PB_{m}$.
If $0< i\leq m$, we assume that we already know
the $\Psi_{0},\Psi_{1},\dots,\Psi_{i-1}$ components of the error word, i.e.,
$\hat{e}_{P}=e_{P}$ for all $P\in\bigcup_{j=0}^{i-1}\Psi_{j}$.
We set a modified received word $(r^{(i)}_{P})_{P\in\PB_m}$
by removing the $\Psi_{0},\Psi_{1},\dots,\Psi_{i-1}$ components
of the error word, i.e.,
\begin{align}\label{eq-cP+e}
r^{(i)}_{P}=
\begin{cases}
r_{P}-e_{P}&{\rm if}\ P\in\bigcup_{j=0}^{i-1}\Psi_{j},\\
r_{P}&{\rm if}\ P\in\bigcup_{j=i}^{m}\Psi_{j}.
\end{cases}
\end{align}
Then, $r^{(i)}_{P}=c_{P}$ if $P\in\bigcup_{j=0}^{i-1}\Psi_{j}$,
and $r^{(i)}_{P}=c_{P}+e_{P}$ if $P\in\bigcup_{j=i}^{m}\Psi_{j}$.

In Step 2, we calculate $S^{(i)}_{h}=\sum_{P\in\PB_{m}}r^{(i)}_{P}h(P)$
for $h\in B_{i}$.
Since $(h(P))_{P\in\PB_{m}}\in\PRM_{\nu}(m,q)^{\perp}$ for $h\in B_{i}$
by Theorem \ref{PRMperp}, we have that
\begin{equation}\label{eq-h(P)inPRMperp}
\sum_{P\in\PB_{m}}c_{P}h(P)=0\quad{\rm for}\ h\in B_{i}.
\end{equation}
\begin{Lem}\label{lem-synd-PRMkaraRM}
We have that $\left(S^{(i)}_{h}\right)_{h\in B_{i}}$ is
the syndrome of $(e_{P})_{P\in\Psi_{i}}$ related to $\RM_{\mu-1}(m-i,q)$, i.e.,
\begin{align}
\left(S^{(i)}_{h}\right)_{h\in B_{i}}
=\left(\sum_{P\in\Psi_{i}}e_{P}h(P)\right)_{h\in B_{i}}\label{thm-syndrome}.
\end{align}
\end{Lem}

\noindent
{\bf Proof:}
Let $h\in B_{i}$.
It follows from Eq.\ \eqref{eq-cP+e} and Eq.\ \eqref{eq-h(P)inPRMperp} that
\begin{align}
S^{(i)}_{h}&=\sum_{P\in\PB_{m}}r^{(i)}_{P}h(P)\\
&=\sum_{P\in\PB_{m}}c_{P}h(P)+\sum_{P\in\bigcup_{j=i}^{m}\Psi_{j}}e_{P}h(P)\\
&=\sum_{P\in\bigcup_{j=i}^{m}\Psi_{j}}e_{P}h(P)
=\sum_{P\in\Psi_{i}}e_{P}h(P),
\end{align}
where $h(P)=0$ for $P\in\bigcup_{j=i+1}^{m}\Psi_{j}$,
since the $i$-th exponent of $h$ is positive
and the $i$-th entry of $P$ is $0$.\hfill$\Box$

In Step 3, if $0\leq i<i_{0}$, then we apply Algorithm \ref{AffineAlgo}
from Eq.\ \eqref{thm-syndrome} as $\Psi=\Psi_{i}$, $B=B_{i}$.
Thus, we obtain the $\Psi_{i}$-component of the error word.
If $i_{0}\leq i\leq m$,
we obtain the $\Psi_{i}$-component of the error word
by applying the IDFT to Eq.\ \eqref{thm-syndrome}.
By repeating Steps 1, 2 and 3 for $i\in\{0,1,\dots,m\}$,
we complete the decoding procedure.
We remark that corresponding codes
to which we apply Algorithm \ref{AffineAlgo} 
are listed in the middle column of Table \ref{Table-correctable}.

\section{Number of errors correctable}\label{Sec-number}
Let $0<\nu\leq m(q-1)$ and $\mu=m(q-1)-\nu$.
Let $\Psi$ be $\PB_{m}$ (or resp. $\Psi_{i}$).
The number of errors correctable for $\PRM_{\nu}(m,q)$
(or resp. $\RM_{\mu-1}(m-i,q)^{\perp}$) is defined by
\begin{equation}
\max\left\{|\Phi|\,\middle|\,\Phi\subseteq\Psi,\ \ 
\begin{array}{c}
(\hat{e}_{P})_{P\in\Psi}=(e_{P})_{P\in\Psi}\\
{\rm for}\ (e_{P})_{P\in\Psi}\in\FB_{q}^{\Psi}\ {\rm with}\\
\Phi=\{P\in\Psi\mid e_{P}\neq 0\}
\end{array}
\right\},
\end{equation}
where $(\hat{e}_{P})_{P\in\Psi}$ is the output of $(e_{P})_{P\in\Psi}$
by applying Algorithm \ref{decodingPRM} to $\PRM_{\nu}(m,q)$
(or resp.\ Algorithm \ref{AffineAlgo} to $\RM_{\mu-1}(m-i,q)^{\perp}$).
We note that the output of $(e_{P})_{P\in\Psi}$ coincides with
that of $(c_{P})_{P\in\Psi}+(e_{P})_{P\in\Psi}$ for all
codewords $(c_{P})_{P\in\Psi}$, since the syndrome does not
depend on codewords.

In this section, we determine the number of errors correctable
for $\PRM_{\nu}(m,q)$.
We recall that Algorithm \ref{decodingPRM} computes an error word correctly
if Algorithm \ref{AffineAlgo} computes the $\Psi_{i}$-component of
the error word correctly for all $i\in\{0,1,\dots,m\}$.
We set
\begin{equation}\label{teiseisuu-Psi0}
t_{0}=\left\lfloor\frac{(q-s)q^{m-r-1}-1}{2}\right\rfloor,
\end{equation}
where $\nu=r(q-1)+s,\ 0\leq s< q-1,\ 0\leq r\leq m-1$.
The numbers of errors correctable for $\RM_{\mu-1}(m-i,q)^{\perp}$
are determined in Proposition \ref{thm-correctable-errors}.
\begin{Prop}\label{thm-correctable-errors}
Let $i_{0}$ be the integer defined in \eqref{eq-def-i_0}.
\begin{enumerate}
\item If $0\leq i< i_{0}$, then the number of errors correctable for $\RM_{\mu-1}(m-i,q)^{\perp}$
is $t_{0}$.\label{number-error-corrctRMdec}
\item If $i_{0}\leq i \leq m$, then the number of errors correctable for $\RM_{\mu-1}(m-i,q)^{\perp}$
is $q^{m-i}$.\label{number-error-corrctIDFT}
\end{enumerate}
\end{Prop}

\noindent
{\bf Proof:}
Assertion \ref{number-error-corrctIDFT} has already been proved.
Here, we prove Assertion \ref{number-error-corrctRMdec}.
Let $0\leq i<i_{0}$.
By \eqref{eq-dual-RM}, we have
\begin{align}
\RM_{\mu-1}(m-i,q)^{\perp}&=\RM_{(m-i)(q-1)-(\mu-1)-1}(m-i,q)\\
&=\RM_{\nu-i(q-1)}(m-i,q).
\end{align}
In addition, by \cite[Proposition 4.16]{Hejinen-Pellikaan},
there exists an ordered basis for $\RM_{\nu-i(q-1)}(m-i,q)$ such that
$d_{{\rm FR}}(\RM_{\nu-i(q-1)}(m-i,q))
=d_{\min}(\RM_{\nu-i(q-1)}(m-i,q))$.
Thus, by \eqref{eq-FR-error}, the number of errors correctable is
\begin{align}
&\left\lfloor\frac{d_{\rm FR}(\RM_{\nu-i(q-1)}(m-i,q))-1}{2}\right\rfloor\\
=&\,\left\lfloor\frac{d_{\min}(\RM_{\nu-i(q-1)}(m-i,q))-1}{2}\right\rfloor\\
=&\,\left\lfloor\frac{(q-s)q^{(m-i)-(r-i)-1}-1}{2}\right\rfloor\quad
({\rm by\ Eq.\ \eqref{eq-MD-RM}})\\
=&\,\left\lfloor\frac{(q-s)q^{m-r-1}-1}{2}\right\rfloor=t_{0}.\quad\Box
\end{align}

The result of Proposition \ref{thm-correctable-errors} is
listed in the rightmost column of Table \ref{Table-correctable}.
\begin{Cor}\label{always-correct}
Let $t$ be the number of errors correctable for $\PRM_{\nu}(m,q)$.
Then, we have $t=t_{0}$.\hfill$\Box$
\end{Cor}

\noindent
{\bf Proof:}
By Theorem \ref{thm-correctable-errors}, we have $t\geq t_{0}$.
If $\{P\in\PB_{m}\mid e_{P}\neq 0\}\subseteq\Psi_{1}$ and
$|\{P\in\PB_{m}\mid e_{P}\neq 0\}|=t_{0}+1$,
it does not always hold that
$(\hat{e}_{P})_{P\in\PB_{m}}=(e_{P})_{P\in\PB_{m}}$.
Hence, we have $t\leq t_{0}$\hfill$\Box$

Thus, the number of errors correctable for $\PRM_{\nu}(m,q)$ is
the same as that for $\RM_{\mu-1}(m,q)^{\perp}$.
In special error cases, Algorithm \ref{decodingPRM}
can correct more errors than $t_{0}$
which is described in Section \ref{Sec-comparison}.

\begin{table}[t]
\caption{(Left) Components of $\PB_{m}$,
(Middle) corresponding codes to which we apply Algorithm \ref{AffineAlgo} and
(Right) component-wise numbers of errors correctable}
\label{Table-correctable}
{\footnotesize
\begin{center}
\begin{tabular}{|c|c|c|}
\hline
Components & Corresponding codes &Numbers of errors correctable\\
\hline
 $\Psi_{0}$ & $\RM_{\mu-1}(m,q)^{\perp}$ & $t_{0}$ \\
 $\Psi_{1}$ & $\RM_{\mu-1}(m-1,q)^{\perp}$ & $t_{0}$ \\
 $\Psi_{2}$ & $\RM_{\mu-1}(m-2,q)^{\perp}$ & $t_{0}$ \\
 $\vdots$ & $\vdots$ & $\vdots$ \\
 $\Psi_{i_{0}-1}$ & $\RM_{\mu-1}(m-i_{0}+1,q)^{\perp}$ & $t_{0}$ \\
 $\Psi_{i_{0}}$ & $\left(\FB_{q}^{\Psi_{m-i_{0}}}\right)^{\perp}$ & $q^{m-i_{0}}=|\Psi_{m-i_{0}}|$ \\
 $\vdots$ & $\vdots$ & $\vdots$ \\
 $\Psi_{m-1}$ & $\left(\FB_{q}^{\Psi_{1}}\right)^{\perp}$ & $q^{1}=|\Psi_{1}|$ \\
 $\Psi_{m}$ & $\left(\FB_{q}^{\Psi_{0}}\right)^{\perp}$ & $1=|\Psi_{0}|$ \\
\hline
\end{tabular}
\end{center}
}
\end{table}

\section{Numerical example}\label{sec-example}
\begin{figure}[h]
{\scriptsize
\begin{center}
\begin{multicols}{2}
Information polynomial $f\in R_{5}$

\vspace{-0.3mm}
$B_{0}$\hspace{-1mm}
{\tabcolsep=1mm
\begin{tabular}{c}
 \\
 \\
$0$\\
$1$\\
$2$\\
$3$\\
\end{tabular}
\hspace{-1.5mm}
\begin{tabular}{@{\vrule width 1.2pt\ }c|c|c|c@{\ \vrule width 1.2pt}@{\ }c|c|c|c@{\ \vrule width 1.2pt}@{\ }c|c|c|c@{\ \vrule width 1.2pt}@{\ }c|c|c|c@{\ \vrule width 1.2pt}}
\multicolumn{4}{c}{$a_{3}=0$}&\multicolumn{4}{c}{$a_{3}=1$}&\multicolumn{4}{c}{$a_{3}=2$}&\multicolumn{4}{c}{$a_{3}=3$}\\
\multicolumn{16}{l}{\hspace{0.5mm}$0$\hspace{2mm}$1$\hspace{2mm}$2$\hspace{2.3mm}$3$\hspace{2.2mm}$0$\hspace{2.1mm}$1$\hspace{2.3mm}$2$\hspace{2.3mm}$3$\hspace{2.2mm}$0$\hspace{2.2mm}$1$\hspace{2.3mm}$2$\hspace{2.3mm}$3$\hspace{2.2mm}$0$\hspace{2.4mm}$1$\hspace{2.2mm}$2$\hspace{2.3mm}$3$}\\
\noalign{\hrule height 1.2pt}
$\beta$&$1$&$\alpha$&$\beta$&$\alpha$&$1$&$0$&$0$&$\beta$&$1$&$\beta$&{\color{white} $\alpha$}&$\beta$&$\beta$&{\color{white} $\alpha$}&{\color{white} $\alpha$}\\
\hline
$1$&$0$&$1$&$1$&$0$&$\alpha$&$\alpha$&{\color{white} $\alpha$}&$0$&$\alpha$&{\color{white} $\alpha$}&{\color{white} $\alpha$}&$0$&{\color{white} $\alpha$}&{\color{white} $\alpha$}&{\color{white} $\alpha$}\\
\hline
$1$&$\beta$&$0$&{\color{white} $\alpha$}&$1$&$0$&{\color{white} $\alpha$}&{\color{white} $\alpha$}&$\beta$&{\color{white} $\alpha$}&{\color{white} $\alpha$}&{\color{white} $\alpha$}&{\color{white} $\alpha$}&{\color{white} $\alpha$}&{\color{white} $\alpha$}&{\color{white} $\alpha$}\\
\hline
$\alpha$&$\alpha$&{\color{white} $\alpha$}&{\color{white} $\alpha$}&$\alpha$&{\color{white} $\alpha$}&{\color{white} $\alpha$}&{\color{white} $\alpha$}&{\color{white} $\alpha$}&{\color{white} $\alpha$}&{\color{white} $\alpha$}&{\color{white} $\alpha$}&{\color{white} $\alpha$}&{\color{white} $\alpha$}&{\color{white} $\alpha$}&{\color{white} $\alpha$}\\
\noalign{\hrule height 1.2pt}
\end{tabular}
}

$B_{1}$\hspace{-1mm}
{\tabcolsep=1mm
\begin{tabular}{c}
 \\
$0$\\
$1$\\
$2$\\
$3$\\
\end{tabular}
\hspace{-1.5mm}
\begin{tabular}{@{\vrule width 1.2pt\ }c|c|c|c@{\ \vrule width 1.2pt}}
\multicolumn{4}{l}{\hspace{0.5mm}$0$\hspace{2.4mm}$1$\hspace{2.2mm}$2$\hspace{2.2mm}$3$}\\
\noalign{\hrule height 1.2pt}
$1$&$\beta$&$0$&$\alpha$\\
\hline
$\alpha$&$\alpha$&$\alpha$&$1$\\
\hline
$\beta$&$\beta$&$0$&{\color{white} $\alpha$}\\
\hline
$\beta$&$0$&{\color{white} $\alpha$}&{\color{white} $\alpha$}\\
\noalign{\hrule height 1.2pt}
\end{tabular}
}
\hspace{2mm}
\ $B_{2}$\hspace{-1mm}
{\tabcolsep=1mm
\begin{tabular}{c}
 \\
$0$\\
$1$\\
$2$\\
$3$\\
\end{tabular}
\hspace{-1.5mm}
\begin{tabular}{@{\vrule width 1.2pt\ }c@{\ \vrule width 1.2pt}}
\multicolumn{1}{l}{ }\\
\noalign{\hrule height 1.2pt}
$\beta$\\
\hline
$0$\\
\hline
$\alpha$\\
\hline
$\alpha$\\
\noalign{\hrule height 1.2pt}
\end{tabular}
}
\hspace{2mm}
\ $B_{3}$\hspace{-0.3mm}
{\tabcolsep=1mm
\begin{tabular}{@{\vrule width 1.2pt\ }c@{\ \vrule width 1.2pt}}
\noalign{\hrule height 1.2pt}
$0$\\
\noalign{\hrule height 1.2pt}
\end{tabular}
}

\rotatebox{270}{\large $\rightarrow$}

Codeword\ $(c_{P})_{P\in\PB_3}$

$\Psi_{0}$\hspace{-1mm}
{\tabcolsep=1mm
\begin{tabular}{c}
 \\
 \\
$0$\\
$1$\\
$\alpha$\\
$\beta$\\
\end{tabular}
\hspace{-1.5mm}
\begin{tabular}{@{\vrule width 1.2pt\ }c|c|c|c@{\ \vrule width 1.2pt}@{\ }c|c|c|c@{\ \vrule width 1.2pt}@{\ }c|c|c|c@{\ \vrule width 1.2pt}@{\ }c|c|c|c@{\ \vrule width 1.2pt}}
\multicolumn{4}{c}{$\omega_{3}=0$}&\multicolumn{4}{c}{$\omega_{3}=1$}&\multicolumn{4}{c}{$\omega_{3}=\alpha$}&\multicolumn{4}{c}{$\omega_{3}=\beta$}\\
\multicolumn{16}{l}{\hspace{0.5mm}$0$\hspace{2.1mm}$1$\hspace{2.1mm}$\alpha$\hspace{2mm}$\beta$\hspace{2mm}$0$\hspace{2.1mm}$1$\hspace{2.3mm}$\alpha$\hspace{2.1mm}$\beta$\hspace{2mm}$0$\hspace{2.2mm}$1$\hspace{2.1mm}$\alpha$\hspace{2.2mm}$\beta$\hspace{2mm}$0$\hspace{2.2mm}$1$\hspace{2.1mm}$\alpha$\hspace{2.2mm}$\beta$}\\
\noalign{\hrule height 1.2pt}
$\beta$&$\alpha$&$0$&$\alpha$&$1$&$\beta$&$1$&$\beta$&$1$&$1$&$\alpha$&$1$&$0$&$0$&$1$&$1$\\
\hline
$\alpha$&$\alpha$&$0$&$0$&$\alpha$&$\beta$&$0$&$0$&$\beta$&$\beta$&$\beta$&$0$&$\alpha$&$\alpha$&$0$&$\alpha$\\
\hline
$0$&$1$&$\beta$&$\alpha$&$1$&$\alpha$&$\alpha$&$\alpha$&$\alpha$&$\beta$&$0$&$0$&$1$&$\alpha$&$0$&$\alpha$\\
\hline
$\alpha$&$\alpha$&$0$&$\beta$&$\alpha$&$1$&$1$&$\beta$&$1$&$1$&$\alpha$&$\beta$&$1$&$0$&$0$&$\alpha$\\
\noalign{\hrule height 1.2pt}
\end{tabular}
}

$\Psi_{1}$\hspace{-1mm}
{\tabcolsep=1mm
\begin{tabular}{c}
 \\
$0$\\
$1$\\
$\alpha$\\
$\beta$\\
\end{tabular}
\hspace{-1.5mm}
\begin{tabular}{@{\vrule width 1.2pt\ }c|c|c|c@{\ \vrule width 1.2pt}}
\multicolumn{4}{l}{\hspace{0.5mm}$0$\hspace{2.1mm}$1$\hspace{2.1mm}$\alpha$\hspace{2mm}$\beta$}\\
\noalign{\hrule height 1.2pt}
$1$&$\alpha$&$0$&$\beta$\\
\hline
$\beta$&$\beta$&$\beta$&$\beta$\\
\hline
$\beta$&$\alpha$&$\alpha$&$0$\\
\hline
$1$&$\beta$&$\beta$&$\alpha$\\
\noalign{\hrule height 1.2pt}
\end{tabular}
}
\hspace{2mm}
\ $\Psi_{2}$\hspace{-1mm}
{\tabcolsep=1mm
\begin{tabular}{c}
 \\
$0$\\
$1$\\
$\alpha$\\
$\beta$\\
\end{tabular}
\hspace{-1.5mm}
\begin{tabular}{@{\vrule width 1.2pt\ }c@{\ \vrule width 1.2pt}}
\multicolumn{1}{l}{ }\\
\noalign{\hrule height 1.2pt}
$\alpha$\\
\hline
$0$\\
\hline
$\beta$\\
\hline
$\beta$\\
\noalign{\hrule height 1.2pt}
\end{tabular}
}
\hspace{2mm}
\ $\Psi_{3}$\hspace{-0.3mm}
{\tabcolsep=1mm
\begin{tabular}{@{\vrule width 1.2pt\ }c@{\ \vrule width 1.2pt}}
\noalign{\hrule height 1.2pt}
$0$\\
\noalign{\hrule height 1.2pt}
\end{tabular}
}

\rotatebox{270}{\large $\rightarrow$}


Received word\ $(r_{P})_{P\in\PB_3}$

$\Psi_{0}$\hspace{-0.3mm}
{\tabcolsep=1mm
\begin{tabular}{@{\vrule width 1.2pt\ }c|c|c|c@{\ \vrule width 1.2pt}@{\ }c|c|c|c@{\ \vrule width 1.2pt}@{\ }c|c|c|c@{\ \vrule width 1.2pt}@{\ }c|c|c|c@{\ \vrule width 1.2pt}}
\noalign{\hrule height 1.2pt}
$\beta$&$\alpha$&$0$&$\alpha$&$1$&$\beta$&$1$&$\beta$&$1$&$1$&$\alpha$&$1$&$0$&$0$&$1$&$1$\\
\hline
$\alpha$&$\alpha$&$0$&$0$&$\alpha$&$\beta$&$0$&$0$&$\beta$&$\beta$&$\beta$&$0$&$\alpha$&$\beta$&$0$&$\alpha$\\
\hline
$0$&$1$&$\beta$&$\alpha$&$1$&$\alpha$&$\alpha$&$\alpha$&$1$&$\beta$&$0$&$0$&$1$&$\alpha$&$0$&$\alpha$\\
\hline
$\alpha$&$0$&$0$&$\beta$&$\alpha$&$1$&$1$&$\beta$&$1$&$1$&$\alpha$&$\beta$&$1$&$0$&$0$&$\alpha$\\
\noalign{\hrule height 1.2pt}
\end{tabular}
}

$\Psi_{1}$\hspace{-0.3mm}
{\tabcolsep=1mm
\begin{tabular}{@{\vrule width 1.2pt\ }c|c|c|c@{\ \vrule width 1.2pt}}
\noalign{\hrule height 1.2pt}
$1$&$\alpha$&$0$&$\beta$\\
\hline
$\beta$&$\beta$&$\beta$&$\alpha$\\
\hline
$\alpha$&$\alpha$&$\alpha$&$0$\\
\hline
$1$&$\beta$&$0$&$\alpha$\\
\noalign{\hrule height 1.2pt}
\end{tabular}
}
\hspace{2mm}
\ $\Psi_{2}$\hspace{-0.3mm}
{\tabcolsep=1mm
\begin{tabular}{@{\vrule width 1.2pt\ }c@{\ \vrule width 1.2pt}}
\noalign{\hrule height 1.2pt}
$\beta$\\
\hline
$\alpha$\\
\hline
$\alpha$\\
\hline
$0$\\
\noalign{\hrule height 1.2pt}
\end{tabular}
}
\hspace{2mm}
\ $\Psi_{3}$\hspace{-0.3mm}
{\tabcolsep=1mm
\begin{tabular}{@{\vrule width 1.2pt\ }c@{\ \vrule width 1.2pt}}
\noalign{\hrule height 1.2pt}
$\beta$\\
\noalign{\hrule height 1.2pt}
\end{tabular}
}

\columnbreak


\rotatebox{270}{\large $\rightarrow$}


Syndrome (black cells)
and its extension in the $\Psi_{0}$-component

$B_{0}$
{\tabcolsep=1mm
\begin{tabular}{@{\vrule width 1.2pt\ }c|c|c|c@{\ \vrule width 1.2pt}@{\ }c|c|c|c@{\ \vrule width 1.2pt}@{\ }c|c|c|c@{\ \vrule width 1.2pt}@{\ }c|c|c|c@{\ \vrule width 1.2pt}}
\noalign{\hrule height 1.2pt}
\cellcolor{black}\color{white}$0$&\cellcolor{black}\color{white}$\beta$&\cellcolor{black}\color{white}${\beta}$&\cellcolor{black}\color{white}${\beta}$&\cellcolor{black}\color{white}${\alpha}$&\cellcolor{black}\color{white}${\beta}$&\cellcolor{black}\color{white}${\beta}$&$\beta$&\cellcolor{black}\color{white}${0}$&\cellcolor{black}\color{white}${\alpha}$&$\alpha$&$\alpha$&\cellcolor{black}\color{white}${\alpha}$&$1$&$1$&$1$\\
\hline
\cellcolor{black}\color{white}${1}$&\cellcolor{black}\color{white}${0}$&\cellcolor{black}\color{white}${0}$&$\alpha$&\cellcolor{black}\color{white}${1}$&\cellcolor{black}\color{white}${\beta}$&$\beta$&$\beta$&\cellcolor{black}\color{white}${1}$&$\alpha$&$\alpha$&$\alpha$&$0$&$1$&$1$&$1$\\
\hline
\cellcolor{black}\color{white}${0}$&\cellcolor{black}\color{white}${\alpha}$&$0$&$\alpha$&\cellcolor{black}\color{white}${0}$&$\beta$&$\beta$&$\beta$&$\beta$&$\alpha$&$\alpha$&$\alpha$&$\beta$&$1$&$1$&$1$\\
\hline
\cellcolor{black}\color{white}${0}$&$\beta$&$0$&$0$&$\alpha$&$\beta$&$\beta$&$\beta$&$0$&$\alpha$&$\alpha$&$\alpha$&$\alpha$&$1$&$1$&$1$\\
\noalign{\hrule height 1.2pt}
\end{tabular}
}

\rotatebox{270}{\large $\rightarrow$}

$\Psi_{0}$-component of the error word

\vspace{-2.5mm}
\rotatebox{270}{\large $\rightarrow$}

Modified received word $(r^{(1)}_{P})_{P\in\PB_3}$

$\Psi_{0}$\hspace{-0.3mm}
{\tabcolsep=1mm
\begin{tabular}{@{\vrule width 1.2pt\ }c|c|c|c@{\ \vrule width 1.2pt}@{\ }c|c|c|c@{\ \vrule width 1.2pt}@{\ }c|c|c|c@{\ \vrule width 1.2pt}@{\ }c|c|c|c@{\ \vrule width 1.2pt}}
\noalign{\hrule height 1.2pt}
$\beta$&$\alpha$&$0$&$\alpha$&$1$&$\beta$&$1$&$\beta$&$1$&$1$&$\alpha$&$1$&$0$&$0$&$1$&$1$\\
\hline
$\alpha$&$\alpha$&$0$&$0$&$\alpha$&$\beta$&$0$&$0$&$\beta$&$\beta$&$\beta$&$0$&$\alpha$&$\alpha$&$0$&$\alpha$\\
\hline
$0$&$1$&$\beta$&$\alpha$&$1$&$\alpha$&$\alpha$&$\alpha$&$\alpha$&$\beta$&$0$&$0$&$1$&$\alpha$&$0$&$\alpha$\\
\hline
$\alpha$&$\alpha$&$0$&$\beta$&$\alpha$&$1$&$1$&$\beta$&$1$&$1$&$\alpha$&$\beta$&$1$&$0$&$0$&$\alpha$\\
\noalign{\hrule height 1.2pt}
\end{tabular}
}

$\Psi_{1}$\hspace{-0.3mm}
{\tabcolsep=1mm
\begin{tabular}{@{\vrule width 1.2pt\ }c|c|c|c@{\ \vrule width 1.2pt}}
\noalign{\hrule height 1.2pt}
$1$&$\alpha$&$0$&$\beta$\\
\hline
$\beta$&$\beta$&$\beta$&$\alpha$\\
\hline
$\alpha$&$\alpha$&$\alpha$&$0$\\
\hline
$1$&$\beta$&$0$&$\alpha$\\
\noalign{\hrule height 1.2pt}
\end{tabular}
}
\hspace{2mm}
\ $\Psi_{2}$\hspace{-0.3mm}
{\tabcolsep=1mm
\begin{tabular}{@{\vrule width 1.2pt\ }c@{\ \vrule width 1.2pt}}
\noalign{\hrule height 1.2pt}
$\beta$\\
\hline
$\alpha$\\
\hline
$\alpha$\\
\hline
$0$\\
\noalign{\hrule height 1.2pt}
\end{tabular}
}
\hspace{2mm}
\ $\Psi_{3}$\hspace{-0.3mm}
{\tabcolsep=1mm
\begin{tabular}{@{\vrule width 1.2pt\ }c@{\ \vrule width 1.2pt}}
\noalign{\hrule height 1.2pt}
$\beta$\\
\noalign{\hrule height 1.2pt}
\end{tabular}
}

\rotatebox{270}{\large $\rightarrow\hspace{-0.5mm}\cdot\cdot\hspace{-0.5mm}\rightarrow$}

Error word $(e_{P})_{P\in\PB_3}$

$\Psi_{0}$\hspace{-0.3mm}
{\tabcolsep=1mm
\begin{tabular}{@{\vrule width 1.2pt\ }c|c|c|c@{\ \vrule width 1.2pt}@{\ }c|c|c|c@{\ \vrule width 1.2pt}@{\ }c|c|c|c@{\ \vrule width 1.2pt}@{\ }c|c|c|c@{\ \vrule width 1.2pt}}
\noalign{\hrule height 1.2pt}
$0$&$0$&$0$&$0$&$0$&$0$&$0$&$0$&$0$&$0$&$0$&$0$&$0$&$0$&$0$&$0$\\
\hline
$0$&$0$&$0$&$0$&$0$&$0$&$0$&$0$&$0$&$0$&$0$&$0$&$0$&$1$&$0$&$0$\\
\hline
$0$&$0$&$0$&$0$&$0$&$0$&$0$&$0$&$0$&$\beta$&$0$&$0$&$0$&$0$&$0$&$0$\\
\hline
$0$&$\alpha$&$0$&$0$&$0$&$0$&$0$&$0$&$0$&$0$&$0$&$0$&$0$&$0$&$0$&$0$\\
\noalign{\hrule height 1.2pt}
\end{tabular}
}

$\Psi_{1}$\hspace{-0.3mm}
{\tabcolsep=1mm
\begin{tabular}{@{\vrule width 1.2pt\ }c|c|c|c@{\ \vrule width 1.2pt}}
\noalign{\hrule height 1.2pt}
$0$&$0$&$0$&$0$\\
\hline
$0$&$0$&$0$&$1$\\
\hline
$1$&$0$&$0$&$0$\\
\hline
$0$&$0$&$\beta$&$0$\\
\noalign{\hrule height 1.2pt}
\end{tabular}
}
\hspace{2mm}
\ $\Psi_{2}$\hspace{-0.3mm}
{\tabcolsep=1mm
\begin{tabular}{@{\vrule width 1.2pt\ }c@{\ \vrule width 1.2pt}}
\noalign{\hrule height 1.2pt}
$1$\\
\hline
$\alpha$\\
\hline
$1$\\
\hline
$\beta$\\
\noalign{\hrule height 1.2pt}
\end{tabular}
}
\hspace{2mm}
\ $\Psi_{3}$\hspace{-0.3mm}
{\tabcolsep=1mm
\begin{tabular}{@{\vrule width 1.2pt\ }c@{\ \vrule width 1.2pt}}
\noalign{\hrule height 1.2pt}
$\beta$\\
\noalign{\hrule height 1.2pt}
\end{tabular}
}
\end{multicols}
\end{center}
}
\caption{Decoding example for $\PRM_{4}(3,4)$}
\label{table-decoding example}
\end{figure}
In this section, we present a numerical example of a decoding procedure
related to a three-dimensional projective space.
To the best of our knowledge, this is the first example
for three-dimensions in the literature.
We consider the case when $m=3,q=4,\nu=5$.
The code length and dimension of $\PRM_{5}(3,4)$ are $n=85$ and $k=50$,
respectively.
By Theorem \ref{PRMperp}, we have $\PRM_{5}(3,4)^{\perp}=\PRM_{4}(3,4)$.
Let $\alpha$ be a generator of a cyclic group $\FB_{4}^{\times}$
satisfying $\alpha^2+\alpha+1=0$, and $\beta$ denotes $\alpha^{2}$.
Then, $\FB_{q}=\{0,1,\alpha,\beta\}$.

Fig.\ \ref{table-decoding example} presents a numerical example
for applying Algorithm \ref{decodingPRM} to $\PRM_{5}(3,4)$.
At Information polynomial of Fig.\ \ref{table-decoding example}, we show
the coefficients of $f\in R_{5}$.
The $(i,j)$th entry of the $4\times4$ matrix named $a_{3}=l$ of $B_{0}$
is the coefficient of
$X_{0}^{5-i-j-l}X_{1}^{i}X_{2}^{j}X_{3}^l$.
Similarly, we show coefficients of $B_{1}$, $B_{2}$ and $B_{3}$ by matricies.
For example, the coefficient of $X_{0}^3 X_{1}^2$ is $\alpha$,
that of $X_{1}^4 X_{2}$ is $\beta$.
At Codeword, we show the values $c_{P}$ indexed by $P\in\PB_{3}$.
For example, $c_{(1:0:1:\beta)}=\alpha$, $c_{(0:0:1:\alpha)}=\beta$.

We have $i_{0}=2$ and $t_{0}=3$.
In the $\Psi_{i}$-component for $i\in\{0,1\}$,
we use the monomial order $\prec$ defined in Section \ref{Sec-RepAVC},
and correct three errors.
For example, if $i=0$, monomials arranged as follows:
$
1\prec X_{1}\prec X_{2}\prec X_{3}\prec X_{1}^2\prec X_{2}X_{1}
\prec X_{2}^2\prec X_{3}X_{1}\prec\cdots.
$
Moreover, we obtain and use Gr{\"o}bner bases
$\GC^{(0)}=\{g_{1}^{(0)}=X_{2}^2+\alpha X_{2}+\beta X_{1},
g_{2}^{(0)}=X_{2}X_{1}+X_{2}+\alpha X_{1}+\alpha,
g_{3}^{(0)}=X_{1}^2+X_{1},
g_{4}^{(0)}=X_{3}+\alpha X_{2}+1\}$ in the $\Psi_{0}$-component,
and 
$\GC^{(1)}=\{g_{1}^{(1)}=X_{3}^2+\beta X_{2}+\beta,
g_{2}^{(1)}=X_{3}X_{2}+X_{3}+\alpha,
g_{3}^{(1)}=X_{2}^2+\beta X_{2}+1\}$ in the $\Psi_{1}$-component.

We correct all error words in the $\Psi_{i}$-component
if $i\in\{2,3\}$.
The number of errors correctable are four and one
in the $\Psi_{2}$- and the $\Psi_{3}$-component, respectively.

\section{Computational complexity}\label{sec-complexity}
In this section,
we calculate computational complexities of Algorithm \ref{decodingPRM}
based on the total number of finite-field operations.
For each $\Psi_{i}$-component of Algorithm \ref{decodingPRM},
the error positions are determined in Step 2 of Algorithm \ref{AffineAlgo} and
the error values $e_{P}$ for all $P\in\Psi_{i}$
are determined in Step 3 of Algorithm \ref{AffineAlgo}.
To observe a precise complexity,
we separate the decoding procedure into the error position determination
and the error value determination.
\begin{Defi}
Let $f(q)$ and $g(q)$ be two functions defined on a subset of
real numbers.
We write $f(q)=O(g(q))$
if and only if there exist constants $q_{0}$ and $C$
such that $|f(q)|\leq C|g(q)|$ for all $q>q_{0}$.\hfill$\Box$
\end{Defi}
Let $N_{i}=q^{m-i}$ be the cardinality of $\Psi_{i}$,
and $z_{i}$ the cardinality of the Gr{\"o}bner basis obtained by
the BMS algorithm for the $\Psi_{i}$-component
for $i\in\{0,1,\dots,m\}$.
\begin{Them}\label{thm-complexity-detection}
Let $n=(q^{m+1}-1)/(q-1)=q^{m}+\cdots+q+1$ the length of $\PRM_{\nu}(m,q)$.
\begin{enumerate}
\item The computational complexity of the error position determination
of Algorithm \ref{decodingPRM} is $O(zn^2)$, where
$z=\max\{z_{0},z_{1},\dots,z_{m}\}\leq N_{0}/q=q^{m-1}<n/q$.
\item The computational complexity of the error value determination
of Algorithm \ref{decodingPRM} is $O(qn^2)$.
\item The total complexity of Algorithm \ref{decodingPRM} is $O(wn^2)$,
where $w=\max\{q, z\}\leq q^{m-1}<n/q$.
\end{enumerate}
\end{Them}
\noindent
{\bf Proof:}
For the $\Psi_{i}$-component,
the computational complexities of the error position determination
and the error value determination are
$O(z_{i}N_{i}^2)=O(z_{i}q^{2m-2i})$ \cite{Cox2}, \cite{Amoros-Sullivan} and
$O(qN_{i}^2)=O(q^{2m-2i+1})$ \cite{Matsuiaffine}, respectively.
According to \cite{Cox2}, \cite{Amoros-Sullivan},
we have $z_{i}\leq N_{i}/q=q^{m-i-1}<N_{i}$ for all $i\in\{0,1,\dots,m\}$.
Hence, the computational complexity of the error position determination
in Algorithm \ref{decodingPRM} is $O(\sum_{i=0}^{m}z_{i}q^{2m-2i})$,
and that of the error value determination is $O(\sum_{i=0}^{m}q^{2m-2i+1})$.

Since the proofs of assertions 1 and 2 are similar
and assertion 3 follows from 1 and 2, we verify only assertion 1.
For all $q>1$, we have $q^2/2<q^2-1$. Thus,
\begin{align}
&z_{0}q^{2m}+z_{1}q^{2m-2}+z_{2}q^{2m-4}+\cdots+z_{m}\label{eq-comp-cal1}\\
\leq \;&z(q^{2m}+q^{2(m-1)}+q^{2(m-2)}+\cdots+1^2)\label{eq-comp-cal2}\\
=\;&z\frac{q^{2m+2}-1}{q^2-1}
<z\frac{2q^{2m+2}}{q^2}=2zq^{2m}.\label{eq-comp-cal3}
\end{align}
This means $\sum_{i=0}^{m}z_{i}q^{2m-2i}=O(zq^{2m})$.
It is clear that $zq^{2m}<zn^2$ for all $q>1$, and then
$zq^{2m}=O(zn^2)$.\hfill$\Box$

We note that Theorem 2 does not depend on $\nu$,
because $\nu$ only affects $|B_{i}|$ which can be replaced by an upper bound
$|\Psi_{i}|=q^{m-i}$ during the complexity analysis.

From the proof of Theorem \ref{thm-complexity-detection},
the computational complexities are $O(yq^{2m})$ and $yq^{2m}=O(yn^2)$,
where $y=z$, $y=q$ or $y=w$.
We also have $yn^2=O(yq^{2m})$.
Indeed, since $(q-1)^2-(q^2/2)=(1/2)(q^2-4q+2)
=(1/2)(q-2)^2-1>0$ for all $q>3$,
we have
\begin{align}
yn^2=y\left(\frac{q^{m+1}-1}{q-1}\right)^2
<y\frac{q^{2m+2}}{(q-1)^2}<2yq^{2m}.
\end{align}
In this sense, Theorem \ref{thm-complexity-detection} is
an optimal evaluation for the computational complexity
of Algorithm \ref{decodingPRM}.

\section{Codeword error rate comparison with MDD}\label{Sec-comparison}
In this section, we investigate the codeword error rate
of Algorithm \ref{decodingPRM} and compare it with that of the MDD
which achieves the best rate of the three previous methods
described in Introduction.
We consider two types of errors correctable.
In the first type, the number of errors correctable is $t_{0}$,
and such errors are always correctable (see Corollary \ref{always-correct}).
The second type is a specialized case,
for which the number of errors correctable has been listed
component-wise in Table \ref{Table-correctable}.
These two types have different codeword error rates.
We refer to the decoding method for the first and second
cases as Proposed Method 1 (PM1) and Proposed Method 2 (PM2), respectively.
Let $p$ be a symbol error rate.
The codeword error rate
of PM1 is then $1-P$, where
$P=\sum_{j=0}^{t_{0}}\binom{n}{j}p^{j}(1-p)^{n-j}$.
The codeword error rate of PM2 is $1-\prod_{i=0}^{i_{0}-1}P_{i}$,
where $P_{i}=\sum_{j=0}^{t_{0}}\binom{q^{m-i}}{j}p^{j}(1-p)^{j}$
for $i\in\{0,1,\dots,i_{0}-1\}$.

Tables \ref{Table-Pro-MDD-num-of-corre1} and \ref{Table-Pro-MDD-num-of-corre2}
list numerical examples
of the number of errors correctable by PM1 and the MDD.
In these tables, the double lines indicate the turning positions of
the quotient obtained when $\nu$ is divided by $q-1$.
The difference between the number of errors correctable decreases
when the above-mentioned quotient increases.
Let $t_{\mathrm{MD}}$ be the number of errors correctable by the MDD.
The codeword error rate of the MDD is
$1-\sum_{j=0}^{t_{\mathrm{MD}}}\binom{n}{j}p^{j}(1-p)^{n-j}
=1-P-\sum_{j=t_{0}+1}^{t_{\mathrm{MD}}}\binom{n}{j}p^{j}(1-p)^{n-j}$.
Recall that $1-P$ is the codeword error rate of PM1.
Therefore, the lower the difference $t_{\mathrm{MD}}-t_{0}$
between the number of errors correctable by PM1 and the MDD,
the lower the difference between their codeword error rates.
In the right hand side of Table \ref{Table-Pro-MDD-num-of-corre1},
i.e., where the quotient obtained by dividing $\nu$ by $q-1$ is $m-1$,
the difference is one or less.
Further, in some cases, the codeword error rate of PM1
coincides with that of the MDD.

\begin{figure}[h]
\begin{center}
\includegraphics[width=90mm]{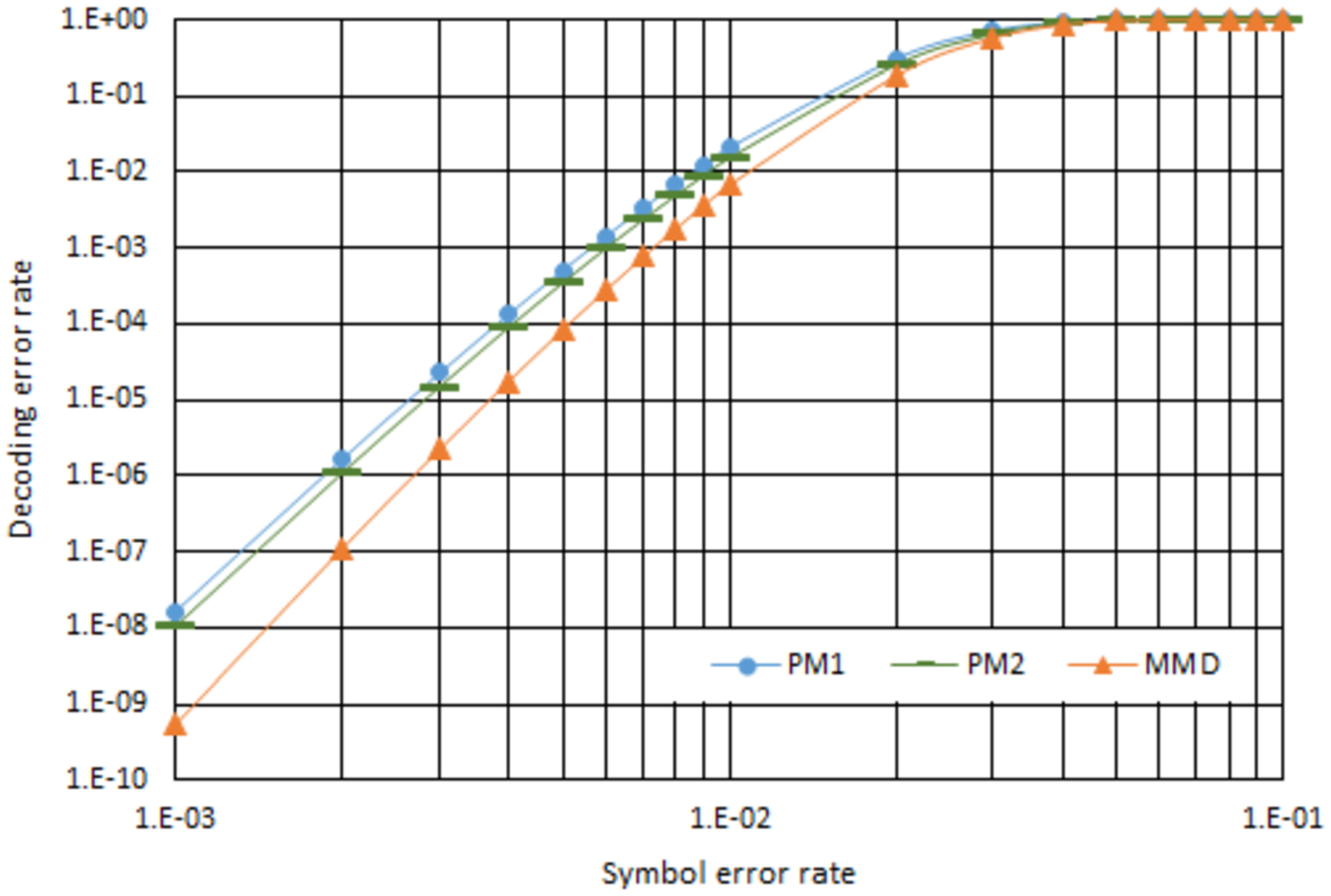}
\end{center}
 \caption{Comparison of codeword error rates for $\PRM_{17}(2,16)$}
 \label{fig-comparison17216}
\begin{center}
\includegraphics[width=90mm]{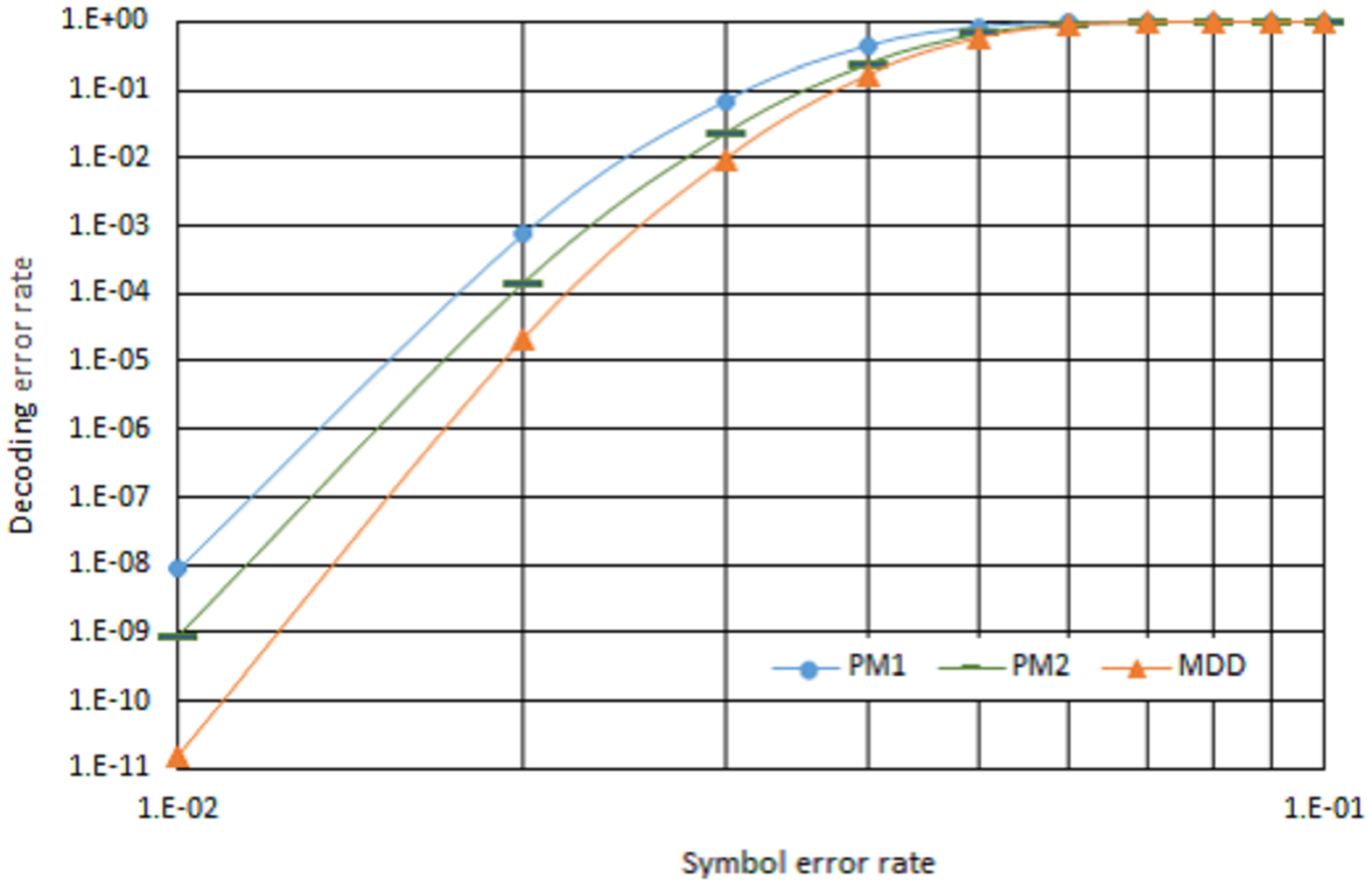}
\end{center}
 \caption{Comparison of codeword error rates for $\PRM_{9}(3,8)$}
 \label{fig-comparison938}
\end{figure}

Figs.\ \ref{fig-comparison17216} and \ref{fig-comparison938} show
the codeword error rates for $\PRM_{17}(2,16)$ and $\PRM_{9}(3,8)$.
When $\nu$ is sufficiently large,
the performance curves of PM1 and PM2 are close to that of the MDD, as
shown in Fig.\ \ref{fig-comparison17216}.
In Fig. \ref{fig-comparison938},
the performance curve of PM2 is distinct from that of PM1
because the cardinality and number of errors correctable are not negligible. 

\begin{table}[t]
{\footnotesize
\caption{Number of errors correctable by
Algorithm\ \ref{decodingPRM} and the MDD for $\PRM_{\nu}(2,16)$}
\label{Table-Pro-MDD-num-of-corre1}
\begin{center}
\begin{tabular}{|c||c|c|c|c||c|c|c|c|c|}
\hline
 $\nu$ &$5$&$8$&$11$&$14$&$17$&$20$&$23$&$26$&$29$\\
\hline
 Algorithm\ \ref{decodingPRM} &$87$&$63$&$39$&$15$&$6$&$5$&$3$&$2$&$0$\\
\hline
 MDD &$95$&$71$&$47$&$23$&$7$&$5$&$4$&$2$&$1$\\
\hline
 Difference &$8$&$8$&$8$&$8$&$1$&$0$&$1$&$0$&$1$\\
\hline
\end{tabular}
\end{center}
\vspace{1mm}
\caption{Number of errors correctable by
Algorithm\ \ref{decodingPRM} and the MDD for $\PRM_{\nu}(3,8)$}
\label{Table-Pro-MDD-num-of-corre2}
\begin{center}
\begin{tabular}{|c||c|c|c||c|c||c|c|c|}
\hline
 $\nu$ &$2$&$4$&$6$&$9$&$12$&$14$&$16$&$18$\\
\hline
 Algorithm \ref{decodingPRM} &$191$&$127$&$63$&$23$&$11$&$7$&$3$&$2$\\
\hline
 MDD &$223$&$159$&$95$&$27$&$15$&$7$&$3$&$2$\\
\hline
 Difference &$32$&$32$&$32$&$4$&$4$&$0$&$0$&$0$\\
\hline
\end{tabular}
\end{center}
}
\end{table}

\section{Conclusion}\label{sec-conclusion}
In this paper, we have constructed a decoding algorithm for all PRM codes by
dividing a projective space into a union of affine spaces.
We have determined the number of errors correctable
for $\PRM_{\nu}(m,q)$.
Although it is the same as the number of errors correctable
for $\RM_{\nu}(m,q)$, advantages of
Algorithm \ref{decodingPRM} are that the codeword is longer and
the code parameters are more flexible.
We have also proved that the computational complexities of
Algorithm \ref{decodingPRM} is $O(wn^2)$,
where $w=\max\{q,z_{0},z_{1},\dots,z_{m}\}$ is less than $n/q$.
Finally, we compared the codeword error rate
of three types of decoding procedures.
When the order of a PRM code is sufficiently high,
the codeword error rate of Algorithm \ref{decodingPRM} is
close to that of the MDD.
Further improvement of our algorithm is required to decrease
the difference between its codeword error rate
and that of the MDD.
This could be a topic for future studies regarding
the decoding theory of PRM codes.
\section*{Acknowledgments}
This work was supported in part by JSPS KAKENHI
Grant Numbers 26887043, 15K13994
and in part by a grant under the Strategic Research Foundation
Grant-aided Project for Private Universities from MEXT (S1311034).


\end{document}